\documentclass[longauth]{aa}
\usepackage{graphicx}
\usepackage[varg]{txfonts}
\usepackage{longtable}
\usepackage[switch]{lineno}
\usepackage{hyperref}
\usepackage{multicol}
\usepackage{color}
\usepackage{natbib}
\bibpunct{(}{)}{;}{a}{}{,} 

\newcommand{\fermi}{\textit{Fermi}}
\newcommand{\fermilat}{\textit{Fermi}-LAT}

\newcommand\eflux{\mbox{${\rm \, erg \,\, cm^{-2} \, s^{-1}}$}}

\begin{document}

   \title{High-energy sources at low radio frequency: the Murchison Widefield Array view of \fermi\ blazars}

   \author{M.\ Giroletti\inst{1}\fnmsep\thanks{Email: giroletti@ira.inaf.it}, F.\ Massaro\inst{2}, R.\ D'Abrusco\inst{3}, R.~Lico\inst{4}, D.\ Burlon\inst{5,6}, N.\ Hurley-Walker\inst{7}, M.\ Johnston-Hollitt\inst{8}, J.~Morgan\inst{7}, V.~Pavlidou\inst{9}, M. Bell\inst{10}, G. Bernardi\inst{11}, R. Bhat\inst{7},
J.D. Bowman\inst{12}, F. Briggs\inst{13,6}, R.J. Cappallo\inst{14}, B.E. Corey\inst{14}, A.A. Deshpande\inst{15}, A. Ewall-Rice\inst{16},  D. Emrich\inst{7}, B.M. Gaensler\inst{5,6,17}, R. Goeke\inst{16}, L.J. Greenhill\inst{18}, B.J. Hazelton\inst{19}, L. Hindson\inst{8}, D.L.Kaplan\inst{20}, J. C.Kasper\inst{21}, E. Kratzenberg\inst{14}, L. Feng\inst{16}, D. Jacobs\inst{12}, N. Kurdryavtseva\inst{7}, E. Lenc\inst{5,6}, C.J. Lonsdale\inst{14}, M. J. Lynch\inst{7}, B. McKinley\inst{22},
S.R. McWhirter\inst{14}, D.A. Mitchell\inst{10,6}, M.F. Morales\inst{19}, E. Morgan\inst{16}, D. Oberoi\inst{23}, A.R. Offringa\inst{24}, S.M. Ord\inst{7,6},
B. Pindor\inst{22}, T. Prabu\inst{15}, P. Procopio\inst{22}, J. Riding\inst{22}, A.E.E. Rogers\inst{14}, A. Roshi\inst{25}, N. Udaya Shankar\inst{15}, K. S. Srivani\inst{15}, R. Subrahmanyan\inst{15,6}, S. J. Tingay\inst{7,6}, M. Waterson\inst{26,7}, R. B. Wayth\inst{7,6}, R. L. Webster\inst{22,6}, A. R. Whitney\inst{14}, A. Williams\inst{7}, C. L. Williams\inst{16}    
           }

   \institute{INAF Osservatorio di Radioastronomia, via Gobetti 101, 40129 Bologna, Italy
\and Physics Department, University of Turin, via Pietro Giuria 1, I-10125 Turin, Italy
\and Department of Physical Sciences, University of Naples Federico II, via Cinthia 9, 80126 Naples, Italy
\and Dipartimento di Fisica e Astronomia, Universit\`a di Bologna, via Ranzani 1, 40127, Bologna, Italy
\and Sydney Institute for Astronomy (SIfA), School of Physics, The University of Sydney, NSW 2006, Australia
\and ARC Centre of Excellence for All-sky Astrophysics (CAASTRO) 
\and International Centre for Radio Astronomy Research (ICRAR), Curtin University, Bentley, WA 6102, Australia
\and School of Chemical and Physical Sciences, Victoria University of Wellington, PO Box 600, Wellington 6140, New Zealand   
\and University of Crete and Foundation for Research and Technology - Hellas, 71003, Heraklion, Greece 
\and CSIRO Astronomy and Space Science (CASS), PO Box 76, Epping, NSW 1710, Australia
\and Department of Physics and Electronics, Rhodes University, PO Box 94, Grahamstown 6140, South Africa
\and School of Earth and Space Exploration, Arizona State University, Tempe, AZ 85287, USA
\and Research School of Astronomy and Astrophysics, Australian National University, Canberra, ACT 2611, Australia 
\and MIT Haystack Observatory, Westford, MA 01886, USA
\and Raman Research Institute, Bangalore 560080, India
\and Kavli Institute for Astrophysics and Space Research, Massachusetts Institute of Technology, Cambridge, MA 02139, USA
\and Dunlap Institute for Astronomy and Astrophysics, University of Toronto, ON, M5S 3H4, Canada
\and Harvard-Smithsonian Center for Astrophysics, Cambridge, MA 02138, USA
\and Department of Physics, University of Washington, Seattle, WA 98195, USA
\and Department of Physics, University of Wisconsin--Milwaukee, Milwaukee, WI 53201, USA
\and Department of Atmospheric, Oceanic and Space Sciences, University of Michigan, Ann Arbor, MI 48109, USA
\and School of Physics, The University of Melbourne, Parkville, VIC 3010, Australia
\and National Centre for Radio Astrophysics, Tata Institute for Fundamental Research, Pune 411007, India
\and Netherlands Institute for Radio Astronomy (ASTRON), PO Box 2, 7990 AA Dwingeloo, The Netherlands
\and National Radio Astronomy Observatory, Charlottesville and Greenbank, USA
\and SKA Organisation, Macclesfield SK11 9DL, UK
}

\date{Received ; accepted }

  \abstract
   {Low-frequency radio arrays are opening a new window for the study of the sky, both to study new phenomena and to better characterize known source classes. Being flat-spectrum sources, blazars are so far poorly studied at low radio frequencies.}
   {We characterize the spectral properties of the blazar population at low radio frequency compare the radio and high-energy properties of the gamma-ray blazar population, and search for radio counterparts of unidentified gamma-ray sources.}
   {We cross-correlated the 6,100 deg$^2$ Murchison Widefield Array Commissioning Survey catalogue with the Roma blazar catalogue, the third catalogue of active galactic nuclei detected by \fermilat, and the unidentified members of the entire third catalogue of gamma-ray sources detected by \fermilat. When available, we also added high-frequency radio data from the Australia Telescope 20 GHz catalogue.}
   {We find low-frequency counterparts for 186 out of 517 (36\%) blazars, 79 out of 174 (45\%) gamma-ray blazars, and 8 out of 73 (11\%) gamma-ray blazar candidates. The mean low-frequency (120--180 MHz) blazar spectral index is $\langle \alpha_\mathrm{low} \rangle=0.57\pm0.02$: blazar spectra are flatter than the rest of the population of low-frequency sources, but are steeper than at $\sim$GHz frequencies. Low-frequency radio flux density and gamma-ray energy flux display a mildly significant and broadly scattered correlation. Ten unidentified gamma-ray sources have a (probably fortuitous) positional match with low radio frequency sources.}
   {Low-frequency radio astronomy provides important information about sources with a flat radio spectrum and high energy. However, the relatively low sensitivity of the present surveys still misses a significant fraction of these objects. Upcoming deeper surveys, such as the GaLactic and Extragalactic All-Sky MWA (GLEAM) survey, will provide further insight into this population.}

\clearpage

   \keywords{BL Lacertae objects: general, catalogues, gamma rays: galaxies, quasars: general, radiation mechanisms: non-thermal, radio continuum: galaxies
               }

  \authorrunning{M.\ Giroletti et al.}
  \titlerunning{High energy sources at low radio frequency: the MWA view of \fermi\ blazars.}

   \maketitle
%

\section{Introduction}

Blazars are the most numerous source population in gamma-ray catalogues \citep{Abdo2010,Nolan2012,Acero2015}. They are radio-loud active galactic nuclei (AGNs) with relativistic jets pointing near to the line of sight. They include flat-spectrum radio quasars (FSRQ), with prominent emission lines in their optical spectra, and BL Lac objects (BL Lacs), with nearly featureless optical spectra. In addition to these markedly different emission line properties, the two classes also have other observational differences; however, the underlying physical processes at work in the two classes are the same, with a beamed relativistic jet powered by accretion onto a supermassive black hole dominating the spectral energy distribution (SED) from radio to gamma rays. In the radio band, at GHz frequencies, blazars of all classes show a flat spectrum ($\alpha<0.5$, in the $S(\nu)\propto\nu^{-\alpha}$ convention). Several recent works demonstrated that blazars largely also maintain a flat spectrum at lower frequency, down to the 300 MHz band \citep{Massaro2013a,Nori2014} and even 74 MHz \citep{Massaro2013b}. However, these studies were based on the comparison of data from low-frequency surveys to $\sim$ GHz surveys carried out at a different epoch. Therefore, they provide only two-point non-simultaneous spectra, which could be affected by time variability (a key feature of blazars), and they are not sensitive to any possible curvature, important information about the physical properties of the emission region, such as the relative contribution of different spectrum components, and breaks in the electron energy distribution.

In the context of the Murchison Widefield Array  \citep[MWA,][]{Tingay2013} instrument commissioning, \citet{Hurley-Walker2014} performed the Murchison Widefield Array Commissioning Survey (MWACS). The MWACS is a multi-wavelength low-frequency radio sky survey, covering approximately 6,100 square degrees in the southern sky over three  bands centred at 119, 150, and 180 MHz. Nearly at the same time, \citet{Massaro2015} published the fifth edition of the Roma-BZCat, the most recent multi-wavelength list of blazars.

At the other end of the electromagnetic spectrum,  \citet{Acero2015} have just released the third \fermilat\ catalogue of gamma-ray sources (3FGL), based on Large Area Telescope (LAT) data collected with a longer exposure and an improved instrument and analysis characterization; furthermore, \citet{Ackermann2015} have realized the accompanying third catalogue of LAT active galactic nuclei (3LAC). These surveys therefore represent an invaluable  resource to tackle the spectral characterization of gamma-ray blazars at low frequency with unprecedented detail and to discuss statistical significance and physical implications of the correlation between emission in the two bands. Indeed, \citet{Ackermann2011a} have studied with great accuracy the radio-gamma connection in the several GHz radio band, while only preliminary studies have been attempted for low-frequency radio data.

In this paper, we cross-correlate the MWACS catalogue with the BZCat, the 3LAC, and the list of unassociated sources of the 3FGL; when available, we also add high-frequency data from the Australia Telescope 20 GHz survey \citep[AT20G,][]{Murphy2010}. All cross-correlations and analyses are carried out within the MWACS footprint, which is given in Sect.~\ref{s.2}, along with an outline of the other radio and gamma-ray surveys and catalogues. Then we describe 
in Sect.~\ref{s.3} the construction of our working samples (blazars in the MWACS, gamma-ray AGNs in the MWACS, other gamma-ray sources in the MWACS) and present their overall properties in Sect.~\ref{s.4}; finally, we discuss the results and give our conclusions in Sect.~\ref{s.5}.

\begin{table}
\centering \small
\caption{List of quantities.  }
\label{t.acronyms}
\begin{tabular}{ll}
\hline
\hline
Symbol & Quantity \\
\hline
$S_{0.18}$ & flux density at 180 MHz (from MWACS) \\
$S_1$ & flux density at $\sim1$ GHz (0.8 GHz from SUMSS or \\ 
 & 1.4 GHz from NVSS, if $\mathrm{Dec.} < -30^\circ$ or $>-30^\circ$,\\
 &  respectively) \\
$S_{20}$ & flux density at 20 GHz (from AT20G) \\
$\alpha_\mathrm{low}$ & spectral index between 120 and 180 MHz (MWACS) \\
$\alpha_{0.18-1}$ & spectral index between 180 MHz and $\sim 1$ GHz \\
$\alpha_{1-20}$ & spectral index between $\sim 1$ and 20 GHz \\
\hline
\end{tabular} 
\end{table}

Throughout the paper, we use a $\Lambda$CDM cosmology with $h = 0.71$, $\Omega_m = 0.27$, and $\Omega_\Lambda=0.73$ \citep{Komatsu2009}.
The radio spectral index $\alpha$ is defined such that $S_\nu \propto \nu^{-\alpha}$ and the gamma-ray  photon index $\Gamma$ such that  $dN_{\rm photon}/dE \propto E^{-\Gamma}$.  In Table~\ref{t.acronyms} we give a list of the quantities used for flux densities and spectral indices throughout the paper.

\section{Selected surveys and catalogues}\label{s.2}


\subsection{MWACS catalogue}

The MWACS catalogue \citep{Hurley-Walker2014} is our main reference catalogue; we downloaded the final table from Vizier\footnote{\url{ftp://cdsarc.u-strasbg.fr/pub/cats/VIII/98}}. This catalogue lists 14,110 sources, in the sky area approximately $0^\circ\le\mathrm{R.A.}\le127.5^\circ$ or $307.5^\circ\le\mathrm{R.A.}\le360^\circ$ ($20.5^\mathrm{h} \le \mathrm{R.A.} \le 8.5^\mathrm{h}$) and $-58.0^\circ < \mathrm{Dec.} < -14.0^\circ$. Data were taken in October 2012.  All the sources in this area have high absolute Galactic latitude, which is ideal for blazar studies (e.g. the 3LAC only contains high-latitude sources by construction).

For each source, the catalogue reports the flux density $S_{0.18}$ at 180 MHz and the spectral index $\alpha_\mathrm{low}$ derived across the three frequency bands centred on 119, 150, and 180 MHz. The survey has $\sim 3\arcmin$ angular resolution and a typical noise level of 40 mJy beam$^{-1}$, with reduced sensitivity near the field boundaries and bright sources. The faintest source has $S_{0.18} = 0.12$ Jy. Sources are marked with a spectral fit flag if the identification at the three frequencies is problematic:\  a type 1 fit classification indicates a spectral index determined by integrating an extended source, and a type 2 a forced fit.

\subsection{BZCat} 

The BZCat is a multi-wavelength list of blazars, regularly updated since the release of the first edition in 2009 \citep{Massaro2009}. \citet{Massaro2015} have recently released the fifth edition, in which they report coordinates, redshift, and multi-frequency (radio, millimetre, optical, X-ray, and gamma-ray) data for 3561 sources. The sources are classified as FSRQs, BL Lacs, and BL Lacs with significant contamination from the host galaxy, or blazars of uncertain type (BCU). All the sources in the BZCat are detected in the radio, and we used the radio flux density at $\sim1$ GHz as reported in the catalogue:\ the radio data reported in the BZCat come partly from the NRAO VLA Sky Survey \citep[NVSS,][]{Condon1998} at 1.4 GHz or the Sydney University Molonglo Sky Survey \citep[SUMSS,][]{Bock1999,Mauch2003} at 0.8 GHz, for sources above or below $\mathrm{Dec.} =-30^\circ$, respectively. In the following, we simply indicate this radio flux density as $S_1$; however, we considered the actual frequency at which the flux density was obtained whenever we used it to determine the spectral index.   

\subsection{3FGL} 

The third \fermilat\ catalogue  (3FGL) is the deepest in the 100 MeV–300 GeV energy range. It is based on the first four years of science operation data from the \fermi\ mission, between 2008 August 4 and 2012 July 31, and it includes 3033 sources. Each source is characterized by its position, gamma-ray flux (photon flux, energy flux, flux in different energy ranges and in 48 monthly time bins), and photon index. The typical 95\% positional confidence radius is  $\sim0.1^{\circ}$.  Most 3FGL sources are identified with or statistically associated with\footnote{An {\it \textup{identification}} is claimed when correlated variability is observed, while the more common {\it \textup{association}} between LAT sources and AGNs is based on statistical methods.} blazars (see Sect.~\ref{s.3lac}), but about one-third of the 3FGL does not have a plausible counterpart at other wavelengths.  These so-called unassociated gamma-ray sources (hereafter, UGS) are not distributed uniformly in the sky: 468 fall in the Galactic plane ($|b|<5^\circ$); they are most likely a mix of galactic and extragalactic discrete sources embedded in complex diffuse emission; the remaining 542 sources are presumably of extragalactic nature, possibly faint and as yet unrecognised blazars. In the present work, we have considered the data available through the \fermi\ Science Support Center\footnote{\url{http://fermi.gsfc.nasa.gov/ssc/data/access/lat/4yr_catalog/gll_psc_v16.fit}}.


\begin{table}
\centering
\caption{Low-frequency detection rates for BZCat and 3LAC source classes in the MWACS footprint.  }
\label{t.rates}
\begin{tabular}{lllll}
\hline
\hline
 & \multicolumn{2}{c}{BZCat} & \multicolumn{2}{c}{3LAC} \\
Class & ratio & \% & ratio & \% \\
\hline
Total & 186/517 & 36\% & 87/247 & 35\% \\
FSRQ & 147/327 & 45\% & 52/71 & 73\% \\
BLL & 23/153 & 15\% & 19/87 & 22\% \\
BCU & 16/37 & 43\% & 8/16 & 50\% \\
Candidates & \dots & \dots & 8/73 & 11\% \\
\hline
\end{tabular} 
\tablefoot{By construction, the BZCat does not contain blazar candidates. For the 3LAC, we did not include radio galaxies.}
\end{table}
\subsection{3LAC \label{s.3lac}}

The third LAT AGN catalogue  (3LAC) includes a total of 1563 gamma-ray sources among the 2192 $|b|>10^\circ$ 3FGL sources. These 3LAC sources are identified or statistically associated with AGNs by means of a Bayesian association \citep{Abdo2010} or a likelihood ratio \citep{Ackermann2011b} method. These 1563 gamma-ray sources are associated with 1591 objects (28 sources have double associations), consisting mostly (98\%) of blazars or blazar candidates.  

From the entire sample, \citet{Ackermann2015} have defined a ``clean'' subset of 3LAC single-association sources free of any analysis problems (e.g.\ sources strongly affected by changes in the diffuse emission model). In the following, whenever we mention the 3LAC we implicitly refer to this clean subset, which includes 1444 objects with 414 FSRQs, 604 BL Lacs, 49 BCUs, 353 blazar candidates, and 24 non-blazar AGNs.  

We note that \citet{Ackermann2015} classify as {\it \textup{blazar candidates}} both confirmed blazars of uncertain type (which have an optical spectrum and are also included in BZCat) and other blazar candidates (which do not appear in BZCat, but still present multi-wavelength features typical of blazars, such as a flat radio spectrum or a two-humped broadband SED). For consistency with the BZCat, we here consider separately gamma-ray blazars of uncertain type (which we indicate as BCU, as in BZCat) and gamma-ray blazar candidates.

In particular, we carried out our analysis starting from the machine-readable form of Table 4 in \citet{Ackermann2015}\footnote{\url{http://iopscience.iop.org/0004-637X/810/1/14/suppdata/apj517471t4_mrt.txt}}.


\subsection{AT20G} 

The Australia Telescope 20 GHz (AT20G) survey is a radio survey carried out at 20 GHz with the Australia Telescope Compact Array (ATCA). It covers the whole sky south of $\mathrm{Dec.}<0^\circ$ and includes 5890 sources above a 20 GHz flux-density limit of 40 mJy. The survey was carried out in two steps from 2004 to 2008: in a first phase, the ATCA realized a fast-scanning blind survey characterized by an overall rms noise of $1\sigma\sim10$ mJy beam$^{-1}$; then, all the sources brighter than 50 mJy were followed up to produce the final catalogue, with additional near-simultaneous flux-density measurements at 5 and 8 GHz for most sources.

The source composition of the AT20G is rather heterogeneous; however, high-frequency observations naturally favour the detection of flat-spectrum sources like blazars and in particular gamma-ray blazars \citep{Mahony2010}. As it also covers the entire MWACS footprint, we used it to complement the MWA data with higher frequency information for the blazar samples that we describe in the following section. In particular, we made use of the data in version 1.0 provided by the Vizier archive\footnote{\url{ftp://cdsarc.u-strasbg.fr/pub/cats/J/MNRAS/402/2403}}.



\section{Sample construction}\label{s.3}

After restricting the BZCat, the 3LAC, and the 3FGL lists to the MWACS footprint, we cross-matched them with the MWACS catalogue using TOPCAT \citep{Taylor2005}. In the following subsections, we give details of the procedure.

\subsection{BZCat-MWACS sample}\label{s.3.1}

Within the entire MWACS field, the BZCat contains 517 sources, divided into 153 BL Lacs, 327 FSRQs, and 37 blazars of uncertain kind. We cross-matched this subset with the MWACS catalogue, using a positional uncertainty of 5\arcsec\ on the BZCat coordinates and the 95\% confidence error ellipse for each MWACS source (obtained as $1.621\times$ the $\mathrm{R.A.}$ and $\mathrm{Dec.}$ $1\sigma$ uncertainty reported in the MWACS catalogue).  The 5\arcsec\ value for the BZCat positions is very conservative for most sources, some of which have positions from VLBI observations that are accurate at the
subarcsecond level. However, we verified that a few additional sources are picked up if we increase the uncertainty from 1\arcsec\ to 5\arcsec, although further increases do not pick up any more sources.

We also tried to cross-match the two catalogues with a fixed radius, using the same procedure based on the generation of 100 mock replicas that we adopted in our previous papers \citep{Massaro2013a,Massaro2013b,Massaro2014}. This method provided a sanity check that the errors quoted in the MWACS catalogue are sensible. Therefore, we continued our analysis using the rigorous association method based on the source-by-source uncertainty. The accuracy of the MWA coordinates is affected by two components, one of random intensity that is due to the ionospheric phase contribution, and one dependent on the source flux density. Since sources with higher flux density have better constrained positions, this method maximizes the efficiency of our selection and the use of information retained from the data.


In total, we found 186 matches: 23 BL Lacs, 147 FSRQs, and 16 blazars of uncertain type. Within these 186 matches, 10 sources have MWACS spectral fit classification of type 1 or 2 (five sources of each type). By constructing 100 mock replicas of the BZCat sample, in which we shifted each position by $2^\circ$ along a random position angle, and repeating the association procedure for each replica, we estimate that less than 1 of these 186 matches arises by chance (on average, 0.8 sources per fake sky).


We furthermore collected high-frequency data for the BZCat-MWACS sample by cross-correlating it with the AT20G survey catalogue. \citet{Murphy2010} calculated the uncertainty in right ascension and declination for the full AT20G sample as $\sigma_\mathrm{RA}=0.9\arcsec$ and $\sigma_\mathrm{Dec} =1.0\arcsec$, respectively. Using these values for AT20G and  extending the positional uncertainty for
BZCat sources from 1\arcsec\ to 5\arcsec, the number of matches increases from 155 to 170, and then does not increase any further up to 30\arcsec.  We then considered all the 170 matches as bona fide associations. The 16 sources that do not have a match are all rather faint at both 180 MHz and 1 GHz, and they also have a steep spectral index ($\langle\alpha_{0.18-1}\rangle=0.68$) and  extrapolated flux densities generally below the AT20G sensitivity.

\subsection{3LAC-MWACS sample}

The subset of the clean 3LAC sources localized within the MWACS footprint contains 249 objects; there are 87 BL Lacs, 71 FSRQ, 16 BCU, 73 blazar candidates, and 2 radio galaxies.

Similar to the BZCat sources, these objects have accurately known coordinates, therefore we followed the same association method described above. We stress that we used the positions of the low-energy counterparts listed in the 3LAC and not those of the gamma-ray sources, which are significantly less well determined  and could result in a large number of spurious associations.


From our starting list, we found counterparts in the MWACS for 88 3LAC sources, divided into 19 BL Lacs, 52 FSRQ, 8 BCU, 8 blazar candidates, and 1 radio galaxy (PKS\,0625$-$35). We note that the other LAT radio galaxy (Pictor A) does indeed appear in the MWACS data as a bright source,  but was not entered  in the final MWACS catalogue because of calibration difficulties brought about by its high brightness and large extent. It therefore does not enter our catalogue cross-correlation either. Given the very low statistics for the population of gamma-ray radio galaxies, in the following we only focus on the 87 3LAC blazars and blazar candidates with counterparts in the MWACS. All but three of the sources (PKS\,0451$-$28, PKS\,2245$-$328, and PKS\,2333$-415$) have spectral fit  class  0. Following the same procedure as
in Sect.~\ref{s.3.1} and considering the smaller number of sources in this case, we do not expect more than one  of these matches to be spurious.

We also cross-matched this sample of 87 sources with the AT20G catalogue and obtained 81 matches. The six missing sources again have fairly steep spectral indices ($\langle\alpha_{0.18-1}\rangle=0.66$) and low extrapolated flux densities (only the blazar candidate PKS\,0302$-$16 exceeds an extrapolated flux density of $S_{20}=100$\,mJy).



\subsection{Other gamma-ray sources in the MWACS}

The MWACS footprint covers 96 3FGL
UGS. These sources have by definition no clearly localized counterpart, therefore we performed the cross correlation of this list with MWACS catalogue using the 95\% confidence radius for each gamma-ray source that is reported in the 3FGL and the positional uncertainty associated with the MWACS coordinates as described in Sect.~\ref{s.3.1}. We found 13 MWACS counterparts for 10 UGS, with seven single matches and three double matches. Three sources have a spectral index flag of 1, and one has 2.

Given the large positional uncertainty of the UGS positions and the local space density of the MWACS sources, we investigated the possibility that these are chance matches. To do this, we created 100 mock replicas of the UGS list, shifting the position by $2^\circ$ in a random position angle, and repeated the association procedure with the same method as in the original list. We find $15\pm3$ MWA matches even starting with a random UGS catalogue. 

For completeness, we also cross-correlated the remaining sources in the 3FGL (i.e.\ those that are neither associated with AGNs nor are UGS), again using the gamma-ray positional uncertainty. We found three more matches:\ two pulsars (PSR\,J0437$-$4715, PSR\,J0742$-$2822) and one starburst galaxy  (NGC\,253).


\begin{figure}
\includegraphics[width=\columnwidth]{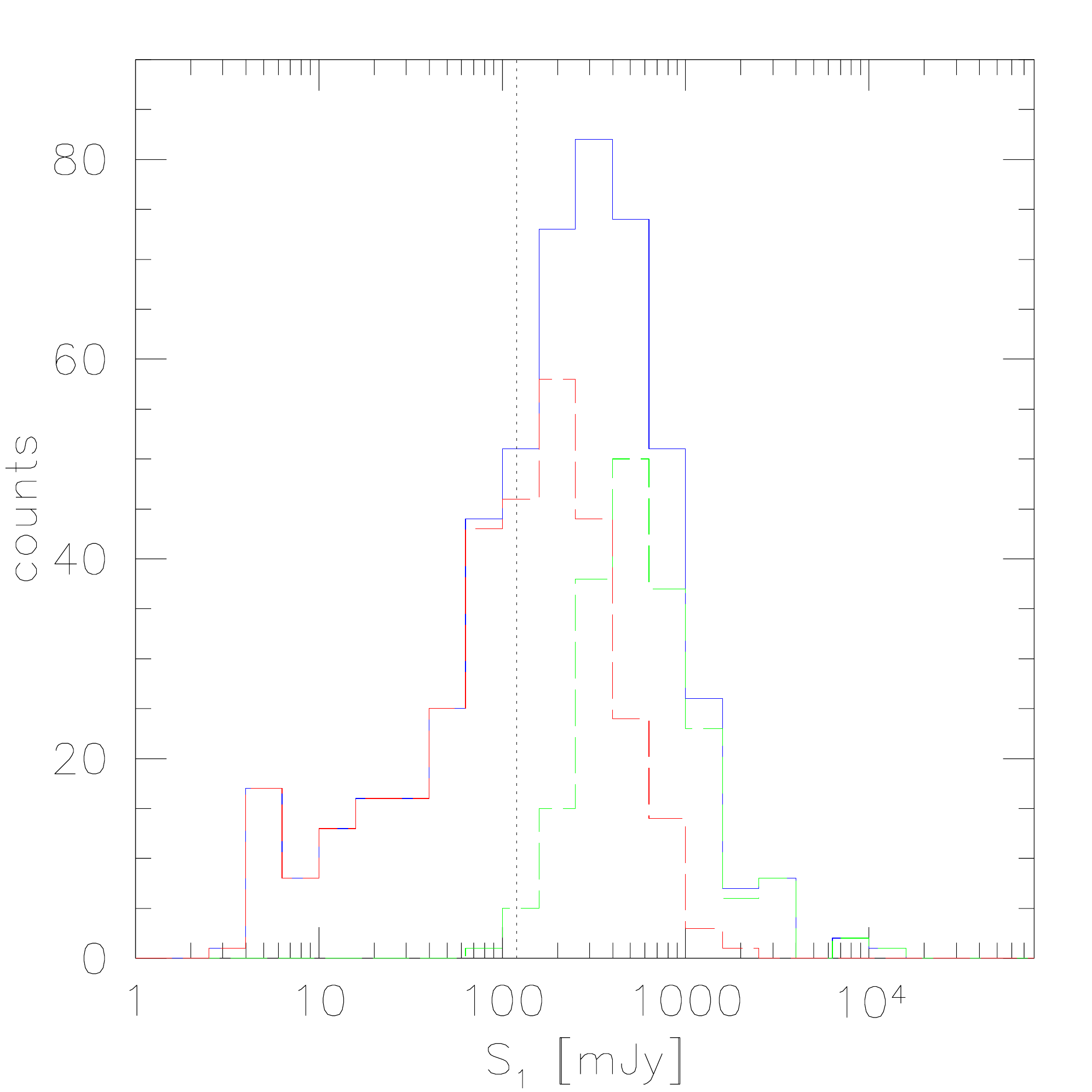}
\caption{Blazar flux density distribution at 1 GHz for the entire BZCat (blue solid line) and for the detected (green dashed line) and not detected (red dashed line) subset in MWACS. The dotted vertical line corresponds to $S_1=120$ mJy, i.e.\ the flux density limit of the MWACS survey extrapolated with $\alpha =0.0$. \label{f.1}}
\end{figure}

\begin{figure*}
\includegraphics[width=\columnwidth]{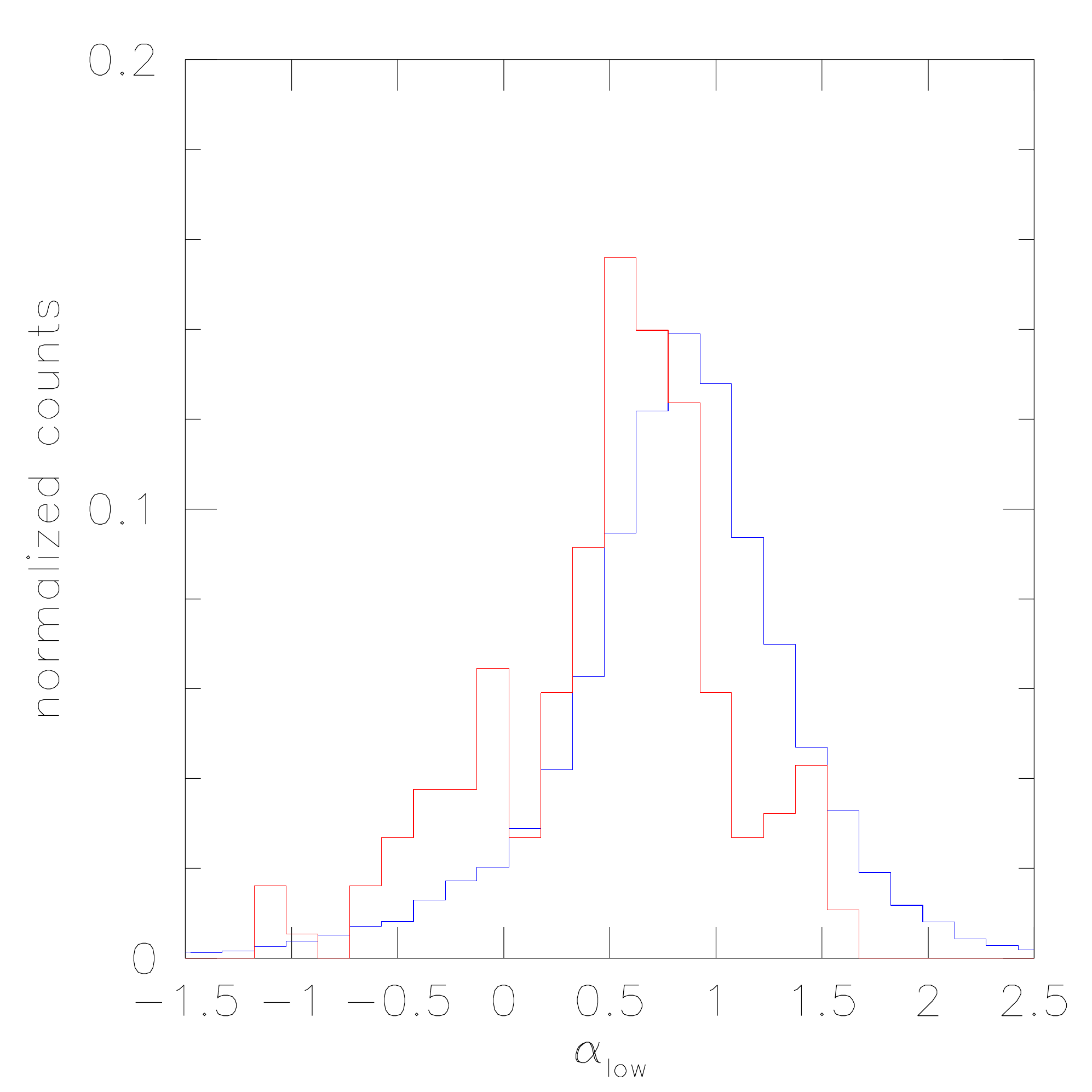}
\includegraphics[width=\columnwidth]{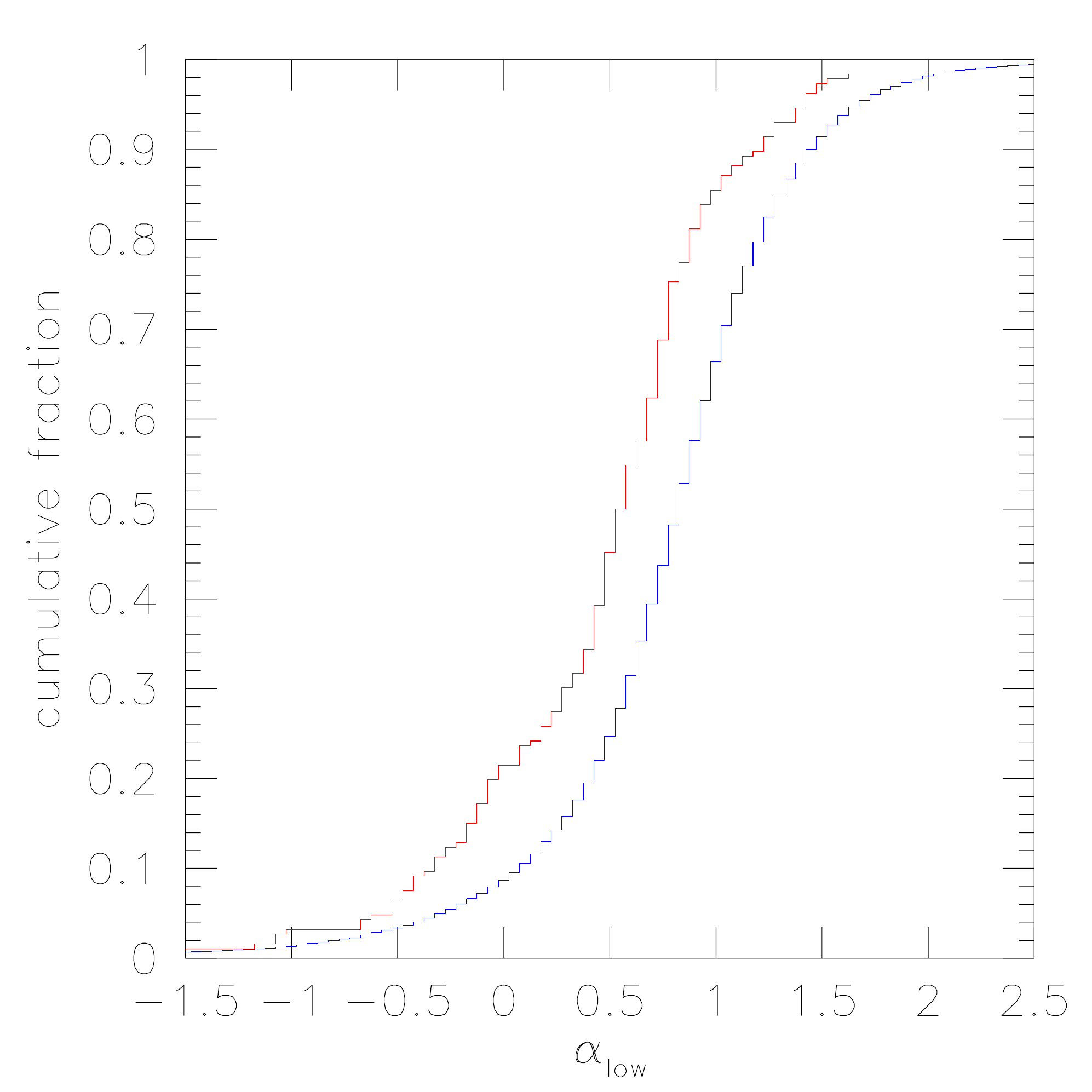}
\caption{Low-frequency spectral index counts (left) and cumulative (right) distributions for the entire MWACS catalogue (blue line) compared with BZCat-MWACS blazars (red line). Because the population counts are very different, normalized distributions are shown. The $x$-axis range is limited to the interval $-1.5<\alpha_\mathrm{low}<2.5$ for illustration purposes. \label{f.2}}
\end{figure*}

\begin{table*}
\centering
\caption{Average spectral indices for BZCat and 3LAC source classes. \label{t.spectra}}
\begin{tabular}{lllllll}
\hline
\hline
Sample & Class & $\langle \alpha_\mathrm{low}\rangle \pm \sigma_{\langle \alpha_\mathrm{low}\rangle}$ & $\langle\alpha_\mathrm{0.18-1}\rangle \pm \sigma_{\langle\alpha_\mathrm{0.18-1}\rangle}$ & $n_\mathrm{low,\ 0.18-1}$ & $\langle\alpha_\mathrm{1-20}\rangle \pm \sigma_{\langle\alpha_\mathrm{1-20}\rangle}$ & $n_{1-20}$ \\
(1) & (2) & (3) & (4) & (5) \\
\hline
BZCat-MWACS & Total & $0.57\pm0.02$ & $0.337\pm0.013$ & 186 & $0.096\pm0.004$ & 170 \\
& FSRQ & $0.56\pm0.02$ & $0.297\pm0.014$ & 147 & $0.097\pm0.005$ & 139 \\
& BLL & $0.49\pm0.05$ & $0.46\pm0.03$ & 23 & $0.105\pm0.014$ & 18 \\
\hline
3LAC-MWACS & Total & $0.50\pm0.03$ & $0.305\pm0.017$ & 87 &$0.074\pm0.005$ & 81 \\
& FSRQ & $0.49\pm0.04$ & $0.18\pm0.02$ & 52 & $0.038\pm0.006$ & 52 \\
& BLL & $0.42\pm0.06$ & $0.39\pm0.04$ & 19 & $0.138\pm0.012$ & 17 \\
\hline
\end{tabular} 
\tablefoot{For each set, we report the weighted average and the error on the weighted average. Weights $w_i$ on each spectral index $\alpha_i$ are determined as $1/\sigma_i$, where $\sigma_i$ is the uncertainty on the spectral index as provided in the MWACS catalogue at low frequency, or is determined through propagation of the uncertainty on the measured flux density for $\alpha_\mathrm{0.18-1}$ and $\alpha_\mathrm{1-20}$. The weighted average is then $\langle \alpha \rangle = \sum_{i=1}^{n} w_i \alpha_i / \sum_{i=1}^{n} w_i$ and the associated error is $\sigma^2_{\langle \alpha \rangle} = 1/ \sum_{i=1}^{n} w_i.$}
\end{table*}

\section{Results}\label{s.4}

Our three catalogues are presented in Tables \ref{t.sample1}--\ref{t.sample3}, that are\ the list of MWACS sources associated with BZCat blazars (Table~\ref{t.sample1}),  with 3LAC gamma-ray blazars (Table~\ref{t.sample2}), and  with un-associated 3FGL gamma-ray sources (Table~\ref{t.sample3}). In the next three subsections, we analyse these catalogues.

\subsection{Blazars in the MWACS: catalogue and demographics}\label{s.4.1}

The 186 matches between BZCat and MWACS listed in Table~\ref{t.sample1} correspond to a detection rate of 36\%. In Cols.\ 2 and 3 of Table 2, we report the detection rates divided by source class, which is clearly higher for FSRQs (45\%) than for BL Lacs (15\%); BCUs appear to have a similar detection rate as FSRQs (albeit with some uncertainty due to the small sample size).

In Fig.~\ref{f.1} we plot the histogram of the $\sim1$ GHz flux density for the entire BZCat sources divided into two subsets according to whether they were detected at low frequency (green histogram) or not (red histogram). This plot  shows that the MWACS detection probability has a strong dependence on the GHz flux density; in other words, the faintest blazars are  very
rarely associated with MWACS counterparts. This immediately explains the highest detection rate found for FSRQs, since FSRQs are on average brighter than BL Lacs. However, it is still remarkable that about a half of the  not detected blazars still have a $\sim1$ GHz flux density higher than that of the faintest source in the MWACS catalogue (167/331 sources have $S_1>120$ mJy). If these sources had  non-inverted spectra ($\alpha_{0.18-1}\ge0.0$), we would expect them to be detected by MWACS, which implies that they have either an inverted spectrum or that they are strongly variable.

In Fig.~\ref{f.2} we also plot the simultaneous low-frequency spectral index $\alpha_\mathrm{low}$ distribution for the entire MWACS catalogue and for the blazars. The cumulative distribution is also shown in the right panel. Even in the 120--180 MHz range, blazars have much flatter spectra than the rest of the radio sources in the extragalactic sky: the weighted average MWACS spectral index for blazars is $\langle \alpha_\mathrm{low} \rangle=0.57\pm0.02$, significantly flatter than the one obtained for the entire MWACS population, for which it is $\langle \alpha_\mathrm{low} \rangle =0.866\pm0.002$. FSRQs are marginally steeper ($\langle \alpha_\mathrm{low} \rangle=0.56\pm0.02$) than BL Lacs ($\langle \alpha_\mathrm{low} \rangle=0.49\pm0.05$). 

For each source, we also calculated the non-simultaneous spectral index $\alpha_\mathrm{0.18-1}$ between 180 MHz and $\sim1$ GHz. The results show a flattening of the spectra from low to higher frequency for FSRQs of about $\langle \Delta\alpha \rangle=\langle \alpha_{0.18-1}\rangle-\langle\alpha_\mathrm{low}\rangle \sim-0.25$, while BL Lacs maintain the same spectral index. This difference between FSRQs and BL Lacs stems from the average low flux density of BL Lacs at $\sim1$ GHz because faint sources can reach the MWACS detection threshold more easily if they steepen at low frequency. 

Finally, for the subset of sources with AT20G counterparts, we also computed the high-frequency spectral index $\langle \alpha_\mathrm{1-20}\rangle$ between $\sim1$ and 20 GHz. In this frequency range, the spectra are much flatter ($\langle \alpha_\mathrm{1-20}\rangle=0.096\pm0.004$), and no difference is found between FSRQs and BL Lacs.
We report all the weighted mean values and the associated errors for the spectral indices in the various frequency ranges in Table \ref{t.spectra}.






\subsection{Gamma-ray blazars in the MWACS: catalogue and demographics}

\begin{figure}
\includegraphics[width=\columnwidth]{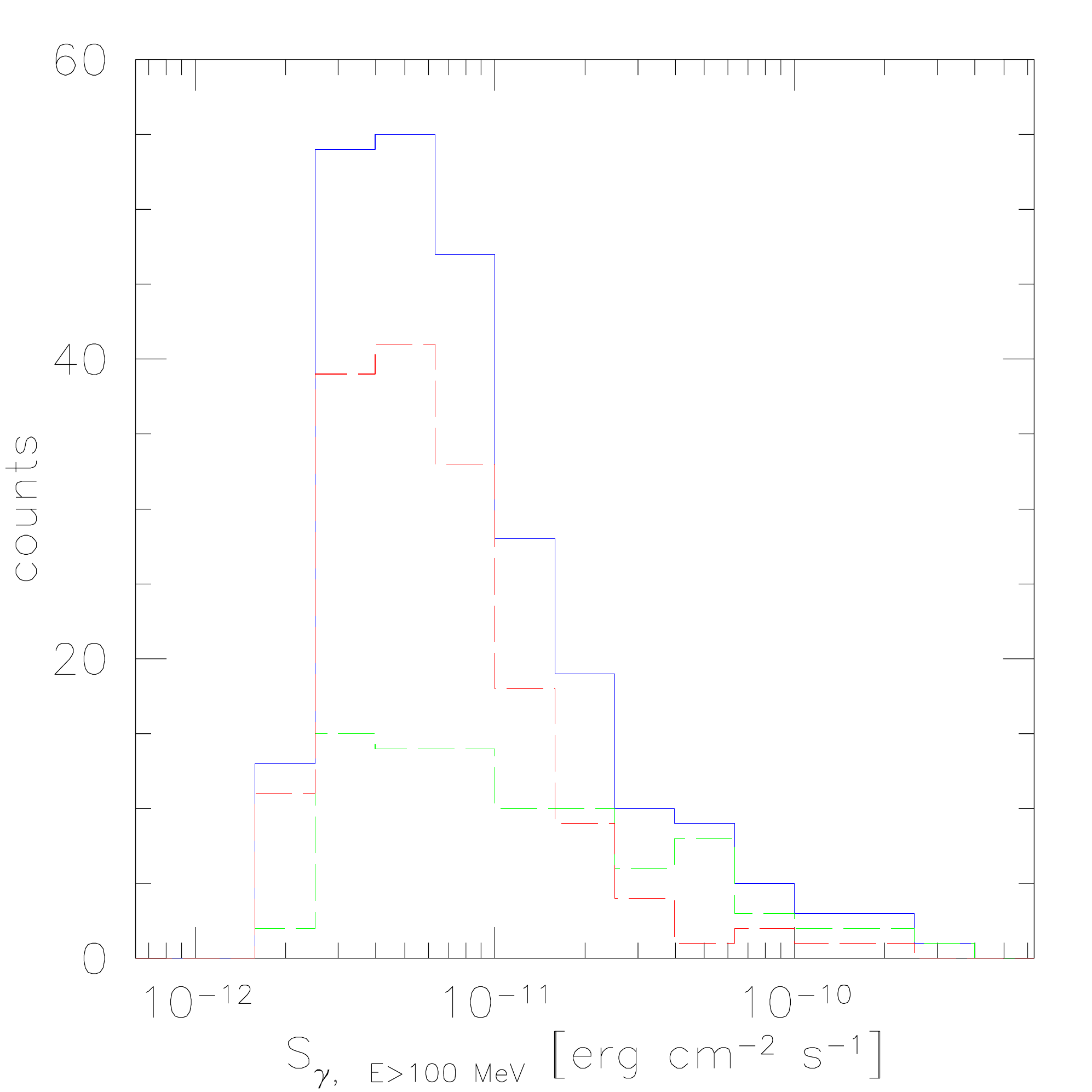}
\caption{Gamma-ray flux at $E>100$ MeV distribution for the entire 3LAC sample (solid blue line). Dashed lines show separately the distribution for 3LAC sources with or without an MWACS counterpart (green and red lines, respectively). \label{f.3}}
\end{figure}

Table \ref{t.sample2} reports the list of the 87 matches between the 3LAC and the MWACS catalogues.  In Table \ref{t.rates} (Cols.\ 4 and 5) we also report the detection rate for the entire sample and for the individual sub-classes. The overall low-frequency detection rate is 35.2\%. If we only consider the confirmed blazars,
however, it increases to 45.4\%, which is significantly higher than for the overall blazar population discussed in Sect.~\ref{s.4.1}. In the sub-classes, the detection rate is highest for FSRQ (73.2\%) and much lower for BL Lacs (21.8\%); BCUs are in between (50.0\%). Again, within each single class, gamma-ray detected sources have a larger detection rate at low frequency than the same kind of sources considered independently of their gamma-ray activity. 
Blazar candidates have a low detection rate of 11.0\%, not surprisingly given their low average 1 GHz radio flux densities.

In Fig.~\ref{f.3} we show the histogram of the gamma-ray energy flux above $E>100$ MeV averaged over the four years of the \fermilat\ data. We indicate separately the set of all the 3LAC sources and the subsets of MWACS detected and not detected sources. The detected sources have higher gamma-ray fluxes or, in other words, the MWACS detection rate is higher for higher gamma-ray fluxes. Moreover, the detection rate above $4\times10^{-11}\eflux$ is 76\% (16/21) and quite uniform across source types. The lower overall MWACS detection rate for BL Lacs with respect to FSRQs arises from the lower gamma-ray flux sources. This indicates that long-term gamma-ray and low-frequency radio fluxes are somehow correlated (see Sect.~\ref{s.4.3}).

\begin{figure}
\includegraphics[width=\columnwidth]{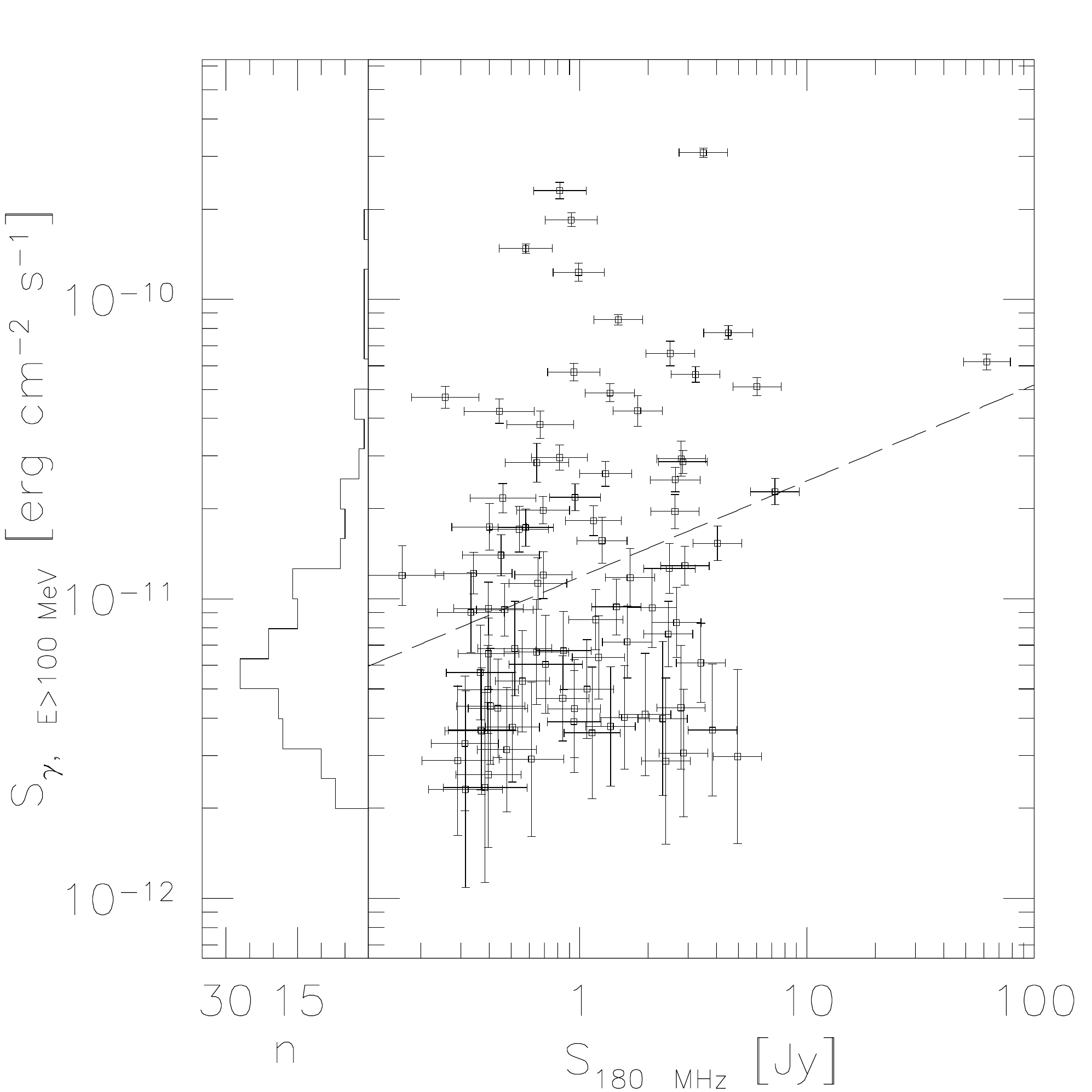}
\caption{\fermilat\ gamma-ray and MWACS fluxes for 3LAC blazars. In the main panel (right), we show the 4 yr gamma-ray energy flux at $E>100$ MeV vs.\ MWACS radio flux density at 180 MHz; the dashed line is the best-fit linear regression. In the smaller left panel, we show the gamma-ray flux distribution for blazars that are not detected in MWACS; gamma-ray flux increases along the $y$-axis, and simple counts increase right-to-left along the $x$-axis. \label{f.radiogamma}}
\end{figure}

The low-frequency spectral index of the gamma-ray blazars is slightly flatter (by about $\Delta \alpha = 0.07$) than the index of the entire blazar population (Table~\ref{t.spectra}, Col.~3). Some flattening is also present in the low-to-mid and mid-to-high spectral indices. For the whole gamma-ray population and within each sub-class, the trend overall is one of flatter spectral indices as higher frequency ranges are considered.

\subsection{Radio and gamma-ray correlation}\label{s.4.3}

\begin{table*}
\centering
\caption{Correlation coefficient and significance for gamma-ray vs.\ MWACS data. \label{t.significance}}
\begin{tabular}{lllllll}
\hline
\hline
Sample & \# objects & \# redshift bins & $r$ & $\rho$ & $p$-value \\
(1) & (2) & (3) & (4) & (5) & (6) \\
\hline
All sources & 87 & -- & 0.26 & 0.27 & -- \\
All sources, with $z$ & 76 & 7 & 0.29 & 0.31 & $6.1\times10^{-2}$ \\
FSRQ & 52 & 5 & 0.21 & 0.25 & 0.15 \\
\hline
\end{tabular} 
\tablefoot{In each row, we indicate the sample considered in Col.\ (1), the number of objects included (Col.\ 2), and the redshift bins used for the statistical analysis (Col.\ 3), the Pearson $r$ and Spearman $\rho$ correlation coefficients and the statistical significance for the correlation between MWACS flux density and gamma-ray energy flux $S_\gamma$ (Cols.\ 4, 5, and 6) determined using the method of \citet{Pavlidou2012}. The method does not consider sources without a measured redshift, therefore values in Cols.\ 3 and 6 are not determined for the entire sample.}
\end{table*}

As suggested in the previous section, it is relevant to search for a possible correlation between low-frequency radio flux densities and gamma-ray fluxes in blazars. We show in Fig.~\ref{f.radiogamma} the scatter plot of the gamma ray energy flux at $E>100$ MeV vs.\ $S_{0.18}$. The $x$-axis range starts at $S_{0.18}= 120$ mJy, which corresponds to the lowest flux density of sources included in the MWACS. Below this threshold, we show the distribution of the gamma-ray flux for the blazars not detected in MWACS.

A simple linear fit yields a slope of $m=0.32\pm0.13$,  a linear correlation coefficient of $r=0.26$, and a null-hypothesis $p$-value of $p=1.5 \times 10^{-2}$, with 87 data points. This fit is somewhat influenced by the brightest radio source (PKS\,0521$-$36). When
we exclude this source from the sample, however, we still obtain a correlation coefficient of $r=0.22$ and a slightly flatter, but still consistent, slope $m=0.30\pm0.14$.  To assess the significance of the observed correlation for the detected sources, we carried out a dedicated analysis based on the method described by \citet{Pavlidou2012} that has been applied by \citet{Ackermann2011a}; this method combines data randomization in luminosity space (to ensure that the randomized data are intrinsically, and not just apparently, uncorrelated) and significance assessment in flux space (to explicitly avoid Malmquist bias and automatically account for the limited dynamic range in both frequencies and the presence of undetected sources). Since the method randomizes luminosities, we considered only sources with a measured redshift, which constitute the large majority of our sample (76 out of 87); moreover, it was shown that for reasonable assumptions on the redshift distribution of the sources without a known $z,$ the method provides a conservative estimate of the significance.

We show the results in Table~\ref{t.significance}. The observed distributions provide evidence of only a low-significance correlation for the entire population ($p=6.1 \times 10^{-2}$), and of no significant correlation at all when only FSRQs are considered ($p=0.15$). We did not consider other smaller sub-samples (e.g. only BL Lacs, or only the flattest spectrum sources) because the number of objects would not have provided a statistically significant result. 

The histogram in the left panel of Fig.~\ref{f.radiogamma} shows that the gamma-ray flux distribution of the gamma-ray blazars without a MWACS counterpart is consistent with the trend estimated for the detected sources. Future, deeper low-frequency surveys will probe this population and provide deeper insight into the possible correlation.

\subsection{Other gamma-ray sources in the MWACS}

We list in Table \ref{t.sample3} the ten UGS that have one or more counterparts in the MWACS catalogue. Given the high spatial density of MWACS sources, it is not possible to claim any of them as a statistically significant association.  We nonetheless list them here as a reference for possible follow-ups. In particular, 3FGL\,J0026.2$-$4812 and 3FGL\,J2130.4$-$4237 are spatially consistent with two moderately bright and flat-spectrum MWACS and SUMSS sources: MWACS\,J0025.6$-$4816, with $S_{0.18}=0.65$ Jy and $\alpha_{0.18-1}=0.50$, and MWACS\,J2131.1-4234, with $S_{0.18}=1.04$ Jy and $\alpha_{0.18-1}=0.63$.

\section{Discussion and conclusions}\label{s.5}

The MWACS is at present the deepest wide-area survey at low frequency. By comparison, the VLA Low-frequency Sky Survey \citep[VLSS][]{Cohen2007} has a typical rms noise level of $\langle \sigma \rangle \approx 0.1$ Jy beam$^{-1}$. The VLSS blazar detection rate reported by \citet{Massaro2013b} was only $\sim26\%$, so that the sample presented in Table \ref{t.sample1} becomes the deepest low-frequency blazar sample ever assembled. Moreover, this sample has simultaneous spectral information by construction, which is extremely valuable for studying core-dominated sources like blazars.

In the widely accepted unified scheme of radio-loud AGNs \citep{Urry1995}, blazars are the aligned counterparts of radio galaxies; in particular, BL Lacs are the aligned counterparts of low-power, edge-dimmed FR1 radio galaxies, and FSRQs are the aligned versions of high radio power, edge-brightened FR2 radio galaxies. This scheme has met with success, and Doppler boosting of radiation emitted from relativistic jets closely aligned to our line of sight (within $\sim 5^\circ$) successfully explains most observational properties of blazars. However, Doppler boosting itself makes the blazar flat-spectrum cores apparently much brighter than the extended radio lobes, which then become very hard to study unless high-sensitivity images with high angular resolution are taken. For this reason, the opening of the low radio frequency window is of great value to the study of the extended emission in blazars \citep{Massaro2013a,Massaro2013b,Massaro2014,Nori2014,Trustedt2014}, and ultimately for the full validation of the unified schemes.
\begin{figure}
\includegraphics[width=\columnwidth]{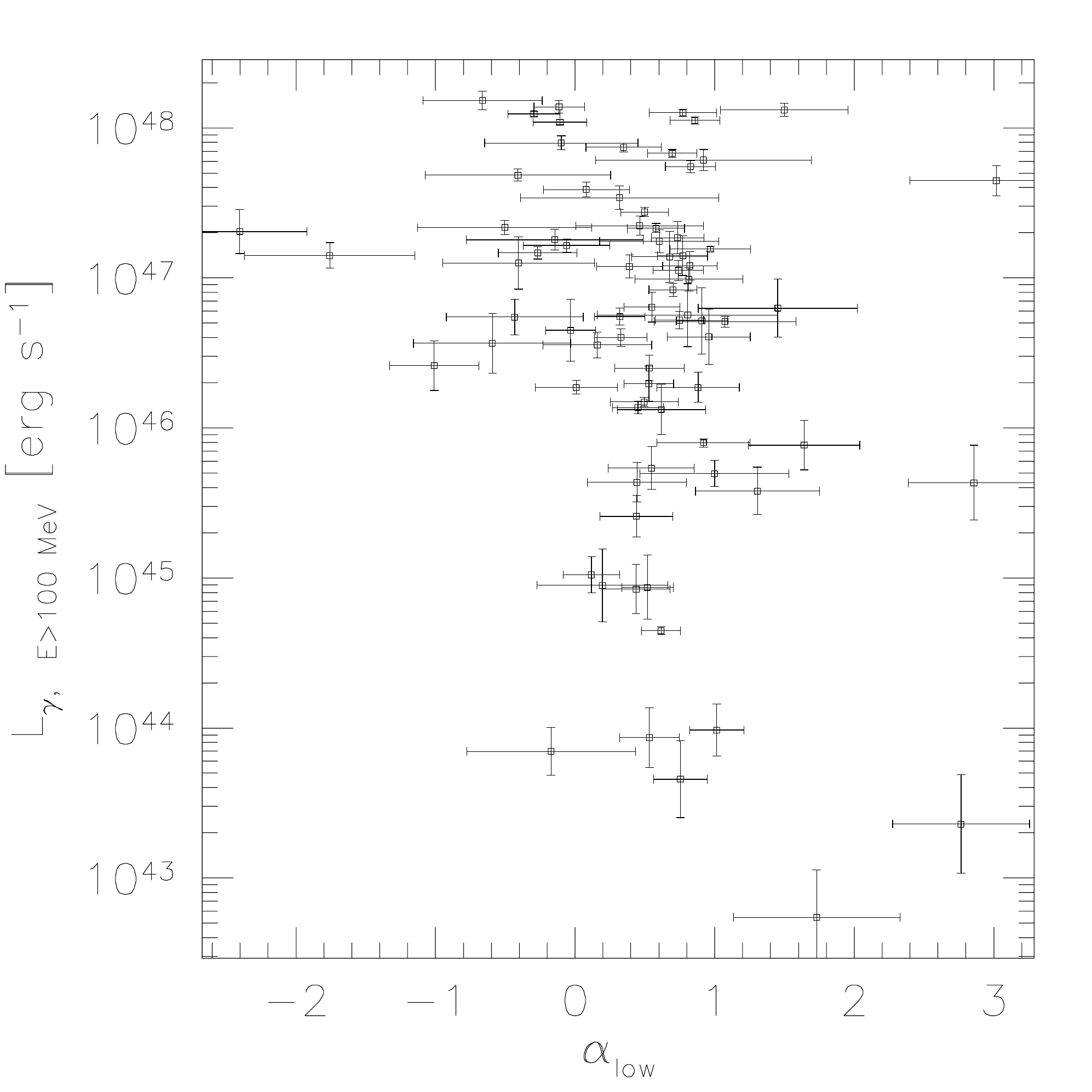}
\caption{Gamma-ray luminosity vs.\ low-frequency spectral index for the 3LAC-MWACS sample.\label{f.lgvsalphalow}}
\end{figure}

Our work has shown that the MWACS spectra of the blazars in our catalogues are on average (1) flatter (by about $\Delta \alpha \sim 0.3$) than those of the entire MWACS population and (2) steeper (by about $\Delta \alpha \sim -0.2$) than those of the same blazar population when considered between 180 MHz and $\sim 1$ GHz. The first fact shows that the core component still contributes significantly to the total emission. The second result is direct proof that extended, steep spectrum emission is also present in blazars.  

The (mildly) significant correlation between the low-frequency flux density and the gamma-ray energy flux, which is produced in the vicinity of the jet base, is also consistent with this scenario. Various works considering higher frequency radio observations have revealed a stronger and more significant correlation between the GHz-domain and gamma-ray data, for instance\ for the 1LAC samples \citep{Ackermann2011a,Mahony2010,Ghirlanda2010,Ghirlanda2011}. The fact that the MWACS data still provide a correlation, but weaker, agrees with additional, but not overwhelming, extended emission that is not beamed and therefore not correlated with the gamma rays.

Our spectral index measurements allowed us to estimate the intensity ratio between the emission from the flat-spectrum core and the steep-spectrum lobes. We assumed that the mean spectral index of the MWACS catalogue describes the extended emission component and that the $\alpha_{1-20}$ measured for our blazar population is a good approximation of the flat spectrum core spectral index. We decomposed the total flux density $S(\nu)$ as
$$S(\nu)=k_c\nu^{-\alpha_c}+k_l\nu^{-\alpha_l} \ ,$$
where $k_c$ and $\alpha_c$ indicate the normalization and spectral index of the core component, and $k_l$ and $\alpha_l$ are the same quantities for the lobes. By substituting $\alpha_c=0.096$ and $\alpha_l=0.866$ and requiring that the average index between 120 and 180 MHz is $\alpha_\mathrm{low}=0.57$, we determined that $k_l/k_c\sim75$.

Although $k_l$ and $k_c$, as well as their simple ratio, have little physical meaning by themselves, they are useful since they allow us to estimate the core-to-lobe flux density ratio at any frequency. For instance, the core-to-lobe flux density ratio is $\sim0.53$ at 120 MHz and $\sim 0.73$ at 180 MHz, indicating that the core still contributes at low frequency, but the majority of the flux density is emitted in the lobes. With increasing frequency, the core becomes the dominant component, with $S_c/S_l\sim3.5$ at 1~GHz, and as large as $\sim27$ at 20 GHz.

Clearly, these values are based on the mean indices and therefore are only suggestive of the average behaviour of the population. Moreover, the MWACS detected sources only constitute about 36\% of the blazar population; about one half of the remaining blazars certainly have flat or inverted spectra, while we can say little about the other half. In general, these results will need to be complemented with deeper low-frequency surveys. Each single source can have wildly different contributions; in particular, the amount of Doppler beaming at the base of the jet can dramatically change the core flux and, accordingly, the component ratio. This is certainly the case for the sources with the flattest (or even inverted) MWACS spectral indices. We can directly test this by comparing the gamma-ray luminosity $L_\gamma$ (which is affected by Doppler beaming) to $\alpha_\mathrm{low}$. In Fig.~\ref{f.lgvsalphalow} we show the plot of these two quantities.  While the points are scattered, the plot shows that the most inverted sources have generally a high gamma-ray luminosity (above $10^{46}$ erg\,s$^{-1}$), whereas the steeper sources span the entire range in luminosity, down to $5.4\times10^{42}$ erg\,s$^{-1}$. Of course, the situation is also complicated by several other factors, such as luminosity distance and source type.

A larger luminosity distance corresponds to higher redshift so that the observed radiation is emitted at higher frequency in the rest frame, where the spectra are flatter; this has a qualitatively similar behaviour to the one produced by the core dominance:\ more luminous sources have flatter spectra both because they are more beamed and because their rest-frame spectra are intrinsically flatter. In this context, we have noted in Sect.~\ref{s.3.1} that most sources in the BZCat-MWACS sample also have a high radio frequency counterpart in AT20G. Of the few that do not, most are typically below the AT20G sensitivity if we extrapolate from low-frequency flux density and $\alpha_{0.18-1}$. Interestingly, there are just two sources that have an extrapolated 20 GHz flux density higher than 100 mJy and they are both FSRQs at high redshift: 5BZQJ2300-2644 ($z=1.476$) and 5BZQJ2327-1447 ($z=2.465$), while the mean $z$ in this sub-sample is $z=0.9$ (in particular, they are the second and fourth most distant in the 16 source subsample).  This could suggest that the rest-frame spectrum of blazars shows a rather complex behaviour, with a new steepening at even higher frequency ($>20$\,GHz) after the flattening between the MWA band and the few GHz domain. A similar high-frequency spectral behaviour for compact sources has been reported by \citet{Chhetri2012} and \citet{Massardi2016}.

We note that the gamma-ray luminosity is generally dependent on the source type \citep{Ackermann2011b,Ackermann2015}, with FSRQs more luminous than BL Lacs. By the peak of the synchrotron component of their SED, blazars are further classified into low-, intermediate-, or high-synchrotron peaked (LSP, ISP, HSP, respectively) sources that are characterized by a synchrotron peak frequency $\nu_\mathrm{peak}$ (Hz) such that $\log \nu_\mathrm{peak}<14$, $14<\log \nu_\mathrm{peak}<15$, and $\log \nu_\mathrm{peak}>15$, respectively. The gamma-ray luminosity decreases from LSP to HSP blazars, similar to what has already been suggested for the blazar sequence by \citet{Fossati1998}. While this scenario is the subject of a lively and long-lasting debate \citep[e.g.][]{Giommi2015,Potter2013,Meyer2011}, we note that our sample is mainly composed of FSRQ (which are typically LSP blazars). The reason for this composition lies in the still somewhat limited sensitivity of the MWACS. As we have noted, the present sample is the deepest available at present; however, faint sources like BL Lacs, and HSP blazars in particular, have lower flux density, and deeper catalogues are necessary to study the blazar population in greater depth.

In the very near future, significant resources are expected to become available, such as the recently released LOw Frequency Array (LOFAR) Multifrequency Snapshot Sky Survey \citep[MSSS, e.g.][]{Heald2015} and the GaLactic and Extragalactic All-Sky MWA Survey \citep[GLEAM,][]{Wayth2015}. On the path towards the Square Kilometer Array, it will be possible to greatly extend the current picture of the connection between high-energy and radio emission in blazars \citep{Giroletti2015} and more generally between the core and extended emission in radio-loud AGNs. The new catalogues will indeed allow us to characterize the low-frequency spectral properties of the still significant population of blazars that are missed in the present work.
MSSS and GLEAM, which also partly overlap, will provide a combined dataset that is ideal for studying the low-frequency properties of \fermilat\ blazars (1) simultaneously and (2) for the full sky. Moreover, because the new surveys will also cover the Galactic plane, important results might be obtained in the study of pulsars and other astrophysical accelerators such as supernovae.

For UGS sources, the situation is more complicated.  The spatial density of low-frequency sources is already too high to claim associations only based on the spatial coincidence, and this is bound to increase at lower flux density limits. It will be necessary to take additional features into account to recognize sources of known gamma-ray emitters. This includes for example a flat spectrum (for core-dominated blazars) or pulsed emission.


\begin{acknowledgements}

MG acknowledges financial support and kind hospitality during his visits at Curtin University and Sydney Institute for Astrophysics. We acknowledge financial contribution from grant PRIN$-$INAF$-$2011. This research has made use of the VizieR catalogue access tool, CDS, Strasbourg, France, and of NASA's Astrophysics Data System.
\\

This scientific work makes use of the Murchison Radio-astronomy Observatory, operated by CSIRO. We acknowledge the Wajarri Yamatji people as the traditional owners of the Observatory site. Support for the operation of the MWA is provided by the Australian Government Department of Industry and Science and Department of Education (National Collaborative Research Infrastructure Strategy: NCRIS), under a contract to Curtin University administered by Astronomy Australia Limited. We acknowledge the iVEC Petabyte Data Store and the Initiative in Innovative Computing and the CUDA Center for Excellence sponsored by NVIDIA at Harvard University. \\

The \textit{Fermi} LAT Collaboration acknowledges generous ongoing support
from a number of agencies and institutes that have supported both the
development and the operation of the LAT  as well as scientific data analysis.
These include the National Aeronautics and Space Administration and the
Department of Energy in the United States, the Commissariat \`a l'Energie Atomique
and the Centre National de la Recherche Scientifique / Institut National de Physique
Nucl\'eaire et de Physique des Particules in France, the Agenzia Spaziale Italiana
and the Istituto Nazionale di Fisica Nucleare in Italy, the Ministry of Education,
Culture, Sports, Science and Technology (MEXT), High Energy Accelerator Research
Organization (KEK) and Japan Aerospace Exploration Agency (JAXA) in Japan, and
the K.~A.~Wallenberg Foundation, the Swedish Research Council and the
Swedish National Space Board in Sweden.
 
Additional support for science analysis during the operations phase is gratefully acknowledged from the Istituto Nazionale di Astrofisica in Italy and the Centre National d'\'Etudes Spatiales in France.

\end{acknowledgements}


\onecolumn
\begin{longtab}{}
\begin{longtable}{lrlrlrrrrr}%
\caption{\label{t.sample1} MWACS-BZCat sources}\\
\hline
  \multicolumn{1}{c}{BZCat name} &
  \multicolumn{1}{c}{Class} &
  \multicolumn{1}{c}{$z$} &
  \multicolumn{1}{c}{$S_1$} &
  \multicolumn{1}{c}{MWACS name} &
  \multicolumn{1}{c}{$S_{0.18}$} &
  \multicolumn{1}{c}{$\sigma_{S_{0.18}}$} &
  \multicolumn{1}{c}{$\alpha_\mathrm{low}$} &
  \multicolumn{1}{c}{$\sigma_{\alpha_\mathrm{low}}$} &
  \multicolumn{1}{c}{Fit} \\
  \multicolumn{1}{c}{} &
  \multicolumn{1}{c}{} &
  \multicolumn{1}{c}{} &
  \multicolumn{1}{c}{(mJy)} &
  \multicolumn{1}{c}{} &
  \multicolumn{1}{c}{(Jy)} &
  \multicolumn{1}{c}{(Jy)} &
  \multicolumn{1}{c}{} &
  \multicolumn{1}{c}{} &
  \multicolumn{1}{c}{} \\
\hline
  5BZQ J0004--4736 & fsrq & 0.884 & 909 & J0004.5--4736 & 0.47 & 0.06 & 0.16 & 0.39 & 0\\
  5BZQ J0005--1648 & fsrq & 0.78 & 263 & J0005.3--1648 & 0.58 & 0.08 & 1.29 & 0.40 & 0\\
  5BZQ J0010--3027 & fsrq & 1.19 & 315 & J0010.5--3027 & 0.45 & 0.06 & 1.42 & 0.34 & 0\\
  5BZQ J0011--2612 & fsrq & 1.096 & 210 & J0011.0--2612 & 0.26 & 0.04 & --0.25 & 0.59 & 0\\
  5BZQ J0015--1812 & fsrq & 0.743 & 387 & J0015.0--1812 & 0.42 & 0.06 & 1.14 & 0.44 & 0\\
  5BZQ J0017--2748 & fsrq & 1.169 & 344 & J0018.0--2748 & 0.24 & 0.04 & --0.50 & 1.02 & 0\\
  5BZQ J0019--3031 & fsrq & 2.677 & 507 & J0019.7--3031 & 0.30 & 0.05 & --0.29 & 0.59 & 0\\
  5BZQ J0025--2227 & fsrq & 0.834 & 202 & J0025.4--2227 & 0.24 & 0.04 & 0.97 & 0.59 & 0\\
  5BZQ J0030--4224 & fsrq & 0.495 & 425 & J0030.3--4224 & 0.69 & 0.08 & 0.01 & 0.29 & 0\\
  5BZQ J0032--2649 & fsrq & 1.47 & 135 & J0032.5--2648 & 0.45 & 0.06 & --0.42 & 0.29 & 1\\
  5BZB J0032--2849 & bll & 0.324 & 160 & J0032.5--2849 & 0.40 & 0.06 & 0.20 & 0.47 & 0\\
  5BZQ J0038--2459 & fsrq & 0.498 & 413 & J0038.2--2459 & 0.40 & 0.05 & --0.43 & 0.49 & 0\\
  5BZB J0040--2719 & bll & 0.172 & 160 & J0040.2--2719 & 0.63 & 0.08 & 0.78 & 0.19 & 1\\
  5BZQ J0049--5738 & fsrq & 1.797 & 2111 & J0050.0--5738 & 3.41 & 0.37 & 0.77 & 0.18 & 0\\
  5BZU J0058--5659 & bcu & \dots & 485 & J0058.7--5658 & 0.37 & 0.06 & 0.96 & 0.50 & 0\\
  5BZQ J0102--2646 & fsrq & 1.597 & 291 & J0102.9--2646 & 0.41 & 0.05 & --0.22 & 0.51 & 0\\
  5BZQ J0115--2804 & fsrq & 2.579 & 439 & J0115.4--2805 & 0.41 & 0.06 & 0.21 & 0.46 & 0\\
  5BZQ J0117--3357 & fsrq & 0.647 & 415 & J0117.7--3357 & 1.48 & 0.16 & 0.69 & 0.20 & 0\\
  5BZQ J0118--2141 & fsrq & 1.165 & 447 & J0118.9--2141 & 0.65 & 0.09 & 0.46 & 0.46 & 0\\
  5BZB J0120--2701 & bll & \dots & 934 & J0120.5--2701 & 1.80 & 0.19 & 0.46 & 0.19 & 0\\
  5BZQ J0124--5113 & fsrq & 1.104 & 251 & J0124.9--5113 & 0.40 & 0.06 & 0.73 & 0.47 & 0\\
  5BZQ J0126--2222 & fsrq & 0.717 & 612 & J0126.2--2222 & 2.67 & 0.29 & 0.53 & 0.18 & 0\\
  5BZQ J0132--1654 & fsrq & 1.02 & 830 & J0132.7--1655 & 1.30 & 0.15 & --0.27 & 0.28 & 0\\
  5BZU J0133--5200 & bcu & \dots & 352 & J0133.1--5200 & 0.31 & 0.05 & 0.46 & 0.64 & 0\\
  5BZQ J0134--3843 & fsrq & 2.14 & 569 & J0134.5--3843 & 0.94 & 0.11 & 0.68 & 0.27 & 0\\
  5BZQ J0135--2008 & fsrq & 1.141 & 559 & J0135.6--2009 & 0.50 & 0.07 & --0.34 & 0.44 & 0\\
  5BZQ J0137--2430 & fsrq & 0.835 & 1181 & J0137.6--2431 & 4.04 & 0.43 & 0.75 & 0.17 & 0\\
  5BZQ J0138--2711 & fsrq & 2.999 & 261 & J0138.1--2711 & 0.35 & 0.05 & 1.26 & 0.51 & 0\\
  5BZQ J0138--2254 & fsrq & 1.895 & 512 & J0138.9--2255 & 2.60 & 0.28 & 0.76 & 0.18 & 0\\
  5BZQ J0145--2733 & fsrq & 1.148 & 923 & J0145.0--2733 & 0.95 & 0.11 & --0.06 & 0.31 & 0\\
  5BZG J0146--5202 & bll-g & 0.098 & 324 & J0146.8--5202 & 1.58 & 0.17 & 1.01 & 0.19 & 0\\
  5BZQ J0153--3310 & fsrq & 0.61 & 1186 & J0153.1--3310 & 0.64 & 0.08 & --0.10 & 0.38 & 0\\
  5BZQ J0154--5107 & fsrq & 1.582 & 447 & J0154.3--5107 & 0.23 & 0.04 & 1.41 & 0.55 & 0\\
  5BZQ J0204--1701 & fsrq & 1.74 & 1220 & J0204.9--1701 & 1.15 & 0.14 & 0.08 & 0.31 & 0\\
  5BZU J0210--5101 & bcu & 1.003 & 3493 & J0210.7--5100 & 6.02 & 0.64 & 0.50 & 0.17 & 0\\
  5BZQ J0222--1615 & fsrq & 0.698 & 590 & J0222.0--1615 & 1.18 & 0.14 & 0.88 & 0.30 & 0\\
  5BZQ J0222--3441 & fsrq & 1.49 & 683 & J0222.9--3441 & 0.44 & 0.07 & --0.63 & 0.61 & 0\\
  5BZQ J0223--5347 & fsrq & 0.569 & 184 & J0223.5--5347 & 0.26 & 0.04 & 1.08 & 0.51 & 0\\
  5BZQ J0228--5546 & fsrq & 2.464 & 352 & J0228.3--5545 & 0.34 & 0.06 & 0.92 & 0.77 & 0\\
  5BZQ J0231--3935 & fsrq & 1.646 & 372 & J0231.8--3935 & 0.41 & 0.06 & 0.00 & 0.53 & 0\\
  5BZQ J0235--4737 & fsrq & 1.504 & 697 & J0235.1--4737 & 0.77 & 0.09 & 0.60 & 0.25 & 0\\
  5BZQ J0236--2953 & fsrq & 2.102 & 313 & J0236.5--2953 & 0.34 & 0.05 & 0.51 & 0.62 & 0\\
  5BZQ J0246--4651 & fsrq & 1.385 & 1498 & J0246.0--4651 & 3.23 & 0.34 & 0.70 & 0.18 & 0\\
  5BZB J0248--1631 & bll & \dots & 520 & J0248.1--1631 & 1.71 & 0.19 & 0.65 & 0.21 & 0\\
  5BZQ J0252--2219 & fsrq & 1.419 & 443 & J0252.8--2219 & 0.94 & 0.11 & 0.35 & 0.27 & 0\\
  5BZQ J0253--5441 & fsrq & 0.539 & 964 & J0253.5--5441 & 0.84 & 0.10 & 0.55 & 0.31 & 0\\
  5BZQ J0256--2137 & fsrq & 1.47 & 367 & J0256.2--2137 & 1.09 & 0.12 & 0.21 & 0.25 & 0\\
  5BZQ J0256--3315 & fsrq & 1.915 & 180 & J0256.7--3315 & 0.37 & 0.06 & 0.91 & 0.62 & 0\\
  5BZQ J0258--5052 & fsrq & 0.834 & 741 & J0258.6--5051 & 1.40 & 0.15 & 0.71 & 0.21 & 0\\
  5BZB J0303--2407 & bll & 0.266 & 700 & J0303.4--2407 & 2.50 & 0.27 & 0.45 & 0.18 & 0\\
  5BZQ J0307--4857 & fsrq & 0.796 & 194 & J0307.6--4856 & 0.40 & 0.05 & 1.43 & 0.41 & 0\\
  5BZQ J0317--2803 & fsrq & 1.166 & 496 & J0317.5--2803 & 1.74 & 0.19 & 0.56 & 0.20 & 0\\
  5BZQ J0321--3122 & fsrq & 1.785 & 322 & J0321.5--3122 & 0.97 & 0.11 & 0.63 & 0.26 & 0\\
  5BZQ J0327--2202 & fsrq & 2.22 & 641 & J0328.0--2202 & 0.82 & 0.10 & --0.04 & 0.31 & 0\\
  5BZQ J0329--2357 & fsrq & 0.895 & 683 & J0329.9--2357 & 2.53 & 0.27 & 0.89 & 0.18 & 0\\
  5BZQ J0331--2524 & fsrq & 2.69 & 320 & J0331.1--2524 & 0.66 & 0.08 & 0.95 & 0.27 & 0\\
  5BZB J0334--4008 & bll & \dots & 1042 & J0334.2--4008 & 0.44 & 0.07 & --0.41 & 0.66 & 0\\
  5BZQ J0336--3616 & fsrq & 1.537 & 501 & J0336.9--3615 & 0.37 & 0.06 & 0.81 & 0.65 & 0\\
  5BZB J0340--2119 & bll & 0.233 & 1075 & J0340.6--2119 & 1.62 & 0.18 & 0.12 & 0.20 & 0\\
  5BZQ J0343--2530 & fsrq & 1.419 & 503 & J0343.3--2530 & 1.45 & 0.16 & 0.82 & 0.19 & 0\\
  5BZQ J0348--2749 & fsrq & 0.991 & 840 & J0348.6--2749 & 1.08 & 0.13 & --1.01 & 0.32 & 0\\
  5BZB J0357--4955 & bll & 0.643 & 215 & J0357.0--4955 & 0.44 & 0.06 & 1.64 & 0.40 & 0\\
  5BZB J0359--2615 & bll & \dots & 796 & J0359.5--2615 & 2.40 & 0.26 & 0.10 & 0.19 & 0\\
  5BZQ J0402--3147 & fsrq & 1.288 & 681 & J0402.3--3147 & 0.37 & 0.05 & --0.59 & 0.56 & 0\\
  5BZQ J0403--2444 & fsrq & 0.598 & 167 & J0403.7--2444 & 0.29 & 0.05 & 2.86 & 0.47 & 0\\
  5BZQ J0403--3605 & fsrq & 1.417 & 1151 & J0403.9--3605 & 1.48 & 0.16 & --0.11 & 0.19 & 0\\
  5BZU J0406--3826 & bcu & 1.285 & 861 & J0407.0--3826 & 0.46 & 0.07 & --0.50 & 0.62 & 0\\
  5BZQ J0411--5149 & fsrq & 1.257 & 351 & J0411.6--5148 & 0.37 & 0.05 & 1.50 & 0.46 & 0\\
  5BZQ J0412--4604 & fsrq & 2.223 & 439 & J0412.8--4604 & 0.23 & 0.04 & 0.48 & 0.87 & 0\\
  5BZU J0416--2056 & bcu & 0.807 & 2779 & J0416.0--2056 & 12.62 & 1.34 & 0.67 & 0.17 & 0\\
  5BZQ J0416--1851 & fsrq & 1.536 & 1248 & J0416.6--1851 & 0.45 & 0.08 & --0.15 & 0.63 & 0\\
  5BZQ J0424--3848 & fsrq & 2.346 & 537 & J0424.5--3848 & 1.47 & 0.16 & 0.40 & 0.22 & 0\\
  5BZQ J0424--3756 & fsrq & 0.782 & 474 & J0424.7--3756 & 0.59 & 0.08 & 0.30 & 0.39 & 0\\
  5BZB J0425--5331 & bll & \dots & 261 & J0425.1--5331 & 0.40 & 0.06 & 1.00 & 0.53 & 0\\
  5BZB J0428--3756 & bll & 1.11 & 753 & J0428.6--3756 & 0.92 & 0.11 & 0.77 & 0.24 & 0\\
  5BZQ J0429--4328 & fsrq & 1.423 & 462 & J0429.4--4328 & 1.40 & 0.15 & 1.00 & 0.21 & 0\\
  5BZQ J0432--5109 & fsrq & 0.557 & 737 & J0432.3--5109 & 1.48 & 0.16 & 0.29 & 0.22 & 0\\
  5BZQ J0434--4355 & fsrq & 2.649 & 474 & J0434.0--4355 & 1.09 & 0.12 & 0.49 & 0.25 & 0\\
  5BZQ J0437--2954 & fsrq & 1.328 & 1092 & J0437.6--2954 & 6.17 & 0.66 & 0.73 & 0.17 & 0\\
  5BZQ J0438--2012 & fsrq & 2.146 & 494 & J0438.8--2012 & 1.37 & 0.15 & 0.40 & 0.22 & 0\\
  5BZU J0439--3210 & bcu & \dots & 374 & J0439.5--3210 & 1.33 & 0.15 & 0.60 & 0.22 & 0\\
  5BZQ J0439--3017 & fsrq & 1.447 & 249 & J0439.9--3017 & 0.32 & 0.05 & 1.55 & 0.47 & 0\\
  5BZQ J0440--4333 & fsrq & 2.852 & 6362 & J0440.3--4333 & 7.63 & 0.81 & --0.05 & 0.17 & 0\\
  5BZQ J0441--4313 & fsrq & 0.593 & 334 & J0441.3--4313 & 0.51 & 0.07 & 1.24 & 0.37 & 0\\
  5BZB J0441--2952 & bll-c & \dots & 576 & J0441.3--2952 & 1.96 & 0.21 & 0.27 & 0.20 & 0\\
  5BZQ J0448--3659 & fsrq & 0.561 & 408 & J0448.2--3659 & 1.61 & 0.17 & 0.73 & 0.17 & 0\\
  5BZB J0449--4350 & bll & 0.205 & 359 & J0449.4--4350 & 0.99 & 0.11 & 0.49 & 0.24 & 0\\
  5BZQ J0451--4653 & fsrq & 0.602 & 476 & J0451.8--4652 & 0.19 & 0.03 & --0.13 & 0.94 & 2\\
  5BZQ J0453--2807 & fsrq & 2.564 & 2541 & J0453.2--2807 & 2.63 & 0.29 & --0.11 & 0.18 & 1\\
  5BZQ J0455--4615 & fsrq & 0.858 & 2100 & J0455.8--4615 & 7.23 & 0.77 & 0.70 & 0.17 & 0\\
  5BZQ J0513--2159 & fsrq & 1.296 & 648 & J0513.8--2159 & 0.52 & 0.07 & 0.12 & 0.40 & 0\\
  5BZQ J0515--4556 & fsrq & 0.194 & 1471 & J0515.7--4556 & 2.37 & 0.26 & 0.82 & 0.22 & 0\\
  5BZQ J0516--1603 & fsrq & 1.278 & 911 & J0516.2--1603 & 3.25 & 0.35 & 0.69 & 0.19 & 0\\
  5BZQ J0521--1737 & fsrq & 0.347 & 459 & J0521.3--1737 & 1.21 & 0.14 & 0.44 & 0.26 & 0\\
  5BZU J0522--3627 & bcu & 0.05655 & 15620 & J0522.9--3627 & 61.86 & 6.39 & 0.62 & 0.14 & 0\\
  5BZQ J0525--4557 & fsrq & 1.479 & 2126 & J0525.5--4557 & 3.85 & 0.41 & 0.91 & 0.18 & 0\\
  5BZQ J0525--4318 & fsrq & 2.164 & 456 & J0525.9--4317 & 1.05 & 0.12 & 1.20 & 0.26 & 0\\
  5BZQ J0526--4830 & fsrq & 1.3 & 377 & J0526.3--4830 & 0.54 & 0.07 & 0.60 & 0.43 & 0\\
  5BZB J0533--4632 & bll & 0.332 & 246 & J0533.7--4631 & 0.67 & 0.09 & 0.78 & 0.40 & 0\\
  5BZQ J0534--3747 & fsrq & 1.668 & 736 & J0534.3--3747 & 2.24 & 0.24 & 0.24 & 0.17 & 0\\
  5BZQ J0536--3401 & fsrq & 0.684 & 652 & J0536.4--3401 & 0.49 & 0.08 & --1.05 & 0.71 & 0\\
  5BZB J0538--4405 & bll & 0.892 & 3729 & J0538.8--4405 & 3.50 & 0.37 & --0.29 & 0.19 & 0\\
  5BZU J0539--1550 & bcu & 0.947 & 599 & J0539.5--1550 & 0.84 & 0.11 & 0.52 & 0.33 & 0\\
  5BZQ J0539--2839 & fsrq & 3.104 & 862 & J0539.9--2839 & 0.58 & 0.07 & --0.66 & 0.43 & 0\\
  5BZQ J0540--5418 & fsrq & 1.185 & 387 & J0540.7--5418 & 0.69 & 0.09 & 0.81 & 0.39 & 0\\
  5BZQ J0549--5246 & fsrq & 0.447 & 140 & J0549.7--5245 & 0.35 & 0.06 & 0.48 & 0.61 & 0\\
  5BZG J0550--3216 & bll-g & 0.069 & 345 & J0550.6--3216 & 2.32 & 0.25 & 0.75 & 0.19 & 0\\
  5BZQ J0559--4529 & fsrq & 0.687 & 384 & J0559.2--4529 & 0.37 & 0.05 & 0.73 & 0.53 & 0\\
  5BZQ J0600--3937 & fsrq & 1.661 & 461 & J0600.5--3937 & 0.64 & 0.09 & --0.41 & 0.54 & 0\\
  5BZQ J0608--2220 & fsrq & 1.926 & 678 & J0609.0--2220 & 0.82 & 0.10 & 0.83 & 0.30 & 0\\
  5BZQ J0609--1542 & fsrq & 0.324 & 2742 & J0609.7--1542 & 2.17 & 0.24 & --1.14 & 0.28 & 0\\
  5BZQ J0612--3138 & fsrq & 0.873 & 613 & J0612.4--3138 & 2.93 & 0.31 & 0.81 & 0.18 & 0\\
  5BZQ J0614--2536 & fsrq & 2.15 & 419 & J0614.2--2537 & 0.65 & 0.08 & --0.14 & 0.40 & 0\\
  5BZU J0620--2515 & bcu & 1.9 & 1214 & J0620.5--2515 & 3.67 & 0.39 & 0.71 & 0.18 & 0\\
  5BZQ J0623--4413 & fsrq & 0.688 & 766 & J0623.5--4413 & 1.40 & 0.15 & 0.58 & 0.21 & 0\\
  5BZB J0629--1959 & bll & \dots & 677 & J0629.4--1959 & 0.67 & 0.10 & --0.10 & 0.55 & 0\\
  5BZQ J0631--4154 & fsrq & 1.416 & 685 & J0631.2--4154 & 1.14 & 0.13 & 0.40 & 0.26 & 0\\
  5BZQ J0632--5404 & fsrq & 0.193 & 171 & J0631.9--5405 & 1.46 & 0.17 & 1.27 & 0.29 & 0\\
  5BZQ J0632--2614 & fsrq & 0.717 & 432 & J0632.1--2614 & 1.55 & 0.17 & 0.59 & 0.21 & 0\\
  5BZQ J0648--4347 & fsrq & 1.029 & 265 & J0648.2--4346 & 1.13 & 0.13 & 1.40 & 0.24 & 0\\
  5BZQ J0648--3044 & fsrq & 1.153 & 898 & J0648.2--3044 & 1.25 & 0.14 & 0.39 & 0.24 & 0\\
  5BZQ J0648--1744 & fsrq & 1.232 & 1046 & J0648.4--1744 & 2.51 & 0.27 & 0.70 & 0.20 & 0\\
  5BZQ J0659--2745 & fsrq & 1.727 & 541 & J0659.8--2745 & 0.60 & 0.07 & 0.45 & 0.37 & 0\\
  5BZQ J0726--4728 & fsrq & 1.686 & 490 & J0726.4--4728 & 0.40 & 0.07 & 0.32 & 0.71 & 0\\
  5BZQ J0728--4745 & fsrq & 2.282 & 376 & J0728.3--4745 & 0.54 & 0.08 & --0.46 & 0.62 & 0\\
  5BZQ J0741--4709 & fsrq & 0.765 & 396 & J0741.7--4709 & 0.29 & 0.05 & 0.27 & 0.89 & 0\\
  5BZQ J2056--4714 & fsrq & 1.491 & 2138 & J2056.2--4714 & 4.50 & 0.48 & 0.86 & 0.18 & 0\\
  5BZU J2105--4848 & bcu & 1.041 & 1100 & J2105.0--4848 & 3.43 & 0.37 & 0.85 & 0.18 & 0\\
  5BZQ J2109--4110 & fsrq & 1.058 & 1823 & J2109.5--4109 & 2.43 & 0.26 & 0.54 & 0.19 & 0\\
  5BZQ J2126--4605 & fsrq & 1.67 & 1520 & J2126.5--4605 & 2.84 & 0.30 & 0.83 & 0.18 & 0\\
  5BZQ J2135--5006 & fsrq & 2.181 & 343 & J2135.3--5006 & 0.17 & 0.03 & 3.02 & 0.62 & 0\\
  5BZQ J2141--3729 & fsrq & 0.423 & 397 & J2141.8--3728 & 0.85 & 0.10 & 0.44 & 0.36 & 0\\
  5BZB J2143--3929 & bll & 0.429 & 143 & J2143.0--3928 & 0.37 & 0.05 & 1.31 & 0.45 & 0\\
  5BZQ J2148--1723 & fsrq & 2.13 & 805 & J2148.6--1723 & 1.85 & 0.22 & 0.59 & 0.28 & 0\\
  5BZQ J2150--2812 & fsrq & 0.865 & 307 & J2150.8--2812 & 0.73 & 0.09 & 0.51 & 0.40 & 0\\
  5BZQ J2151--2742 & fsrq & 1.485 & 315 & J2151.3--2742 & 0.41 & 0.06 & 1.45 & 0.57 & 0\\
  5BZQ J2151--3027 & fsrq & 2.345 & 1244 & J2151.9--3027 & 0.82 & 0.10 & 1.50 & 0.46 & 0\\
  5BZQ J2158--1501 & fsrq & 0.672 & 3021 & J2158.0--1501 & 2.48 & 0.28 & 0.53 & 0.25 & 0\\
  5BZB J2158--3013 & bll & 0.116 & 489 & J2158.8--3013 & 0.82 & 0.10 & 0.92 & 0.33 & 0\\
  5BZQ J2200--1632 & fsrq & 0.836 & 240 & J2200.9--1632 & 0.93 & 0.12 & 0.52 & 0.44 & 0\\
  5BZQ J2206--1835 & fsrq & 0.619 & 6399 & J2206.1--1835 & 10.40 & 1.11 & 0.39 & 0.17 & 0\\
  5BZQ J2207--5346 & fsrq & 1.215 & 1799 & J2207.7--5346 & 2.91 & 0.31 & 0.74 & 0.18 & 0\\
  5BZQ J2207--5707 & fsrq & 2.725 & 474 & J2207.9--5707 & 0.66 & 0.09 & 0.45 & 0.39 & 0\\
  5BZU J2209--2453 & bcu & 0.159 & 455 & J2209.3--2453 & 3.87 & 0.41 & 0.79 & 0.18 & 0\\
  5BZQ J2213--2529 & fsrq & 1.833 & 1210 & J2213.0--2529 & 2.46 & 0.26 & 0.73 & 0.19 & 0\\
  5BZQ J2219--2719 & fsrq & 3.634 & 304 & J2219.5--2718 & 0.31 & 0.05 & --0.29 & 0.79 & 2\\
  5BZQ J2223--3455 & fsrq & 0.298 & 850 & J2223.1--3455 & 2.86 & 0.31 & 0.52 & 0.18 & 0\\
  5BZQ J2230--3942 & fsrq & 0.318 & 370 & J2230.6--3942 & 1.09 & 0.13 & 1.04 & 0.26 & 0\\
  5BZQ J2230--4416 & fsrq & 1.326 & 555 & J2230.9--4416 & 0.51 & 0.06 & 0.96 & 0.30 & 0\\
  5BZQ J2232--1659 & fsrq & 1.78 & 535 & J2232.3--1658 & 0.72 & 0.10 & --0.03 & 0.39 & 1\\
  5BZQ J2239--5525 & fsrq & 1.975 & 296 & J2239.1--5525 & 0.29 & 0.05 & 0.02 & 0.64 & 0\\
  5BZB J2243--2544 & bll & 0.774 & 1102 & J2243.4--2544 & 2.63 & 0.28 & 0.32 & 0.18 & 0\\
  5BZQ J2246--5607 & fsrq & 1.325 & 510 & J2246.3--5607 & 0.73 & 0.09 & 0.48 & 0.30 & 0\\
  5BZQ J2247--3657 & fsrq & 2.252 & 1261 & J2247.0--3657 & 4.68 & 0.48 & 0.67 & 0.14 & 0\\
  5BZQ J2248--2702 & fsrq & 2.75 & 89 & J2248.6--2703 & 0.57 & 0.07 & 1.45 & 0.27 & 0\\
  5BZQ J2248--3235 & fsrq & 2.268 & 708 & J2248.6--3235 & 0.39 & 0.05 & --2.40 & 0.48 & 2\\
  5BZQ J2249--3039 & fsrq & 1.307 & 432 & J2249.3--3039 & 1.40 & 0.15 & 0.80 & 0.20 & 0\\
  5BZU J2250--2806 & bcu & 0.525 & 306 & J2250.7--2806 & 0.26 & 0.04 & 1.07 & 0.51 & 0\\
  5BZU J2254--3209 & bcu & 0.189 & 655 & J2254.5--3209 & 2.94 & 0.31 & 0.74 & 0.18 & 0\\
  5BZQ J2254--4139 & fsrq & 1.765 & 435 & J2254.6--4139 & 0.48 & 0.07 & 1.17 & 0.44 & 0\\
  5BZQ J2256--2735 & fsrq & 1.751 & 425 & J2255.9--2735 & 0.68 & 0.08 & 0.80 & 0.19 & 1\\
  5BZQ J2258--2758 & fsrq & 0.926 & 1249 & J2258.1--2758 & 1.36 & 0.15 & 0.58 & 0.20 & 0\\
  5BZQ J2300--2644 & fsrq & 1.476 & 712 & J2300.4--2644 & 2.12 & 0.23 & 0.64 & 0.18 & 0\\
  5BZU J2303--1841 & bcu & 0.129 & 861 & J2303.0--1841 & 4.91 & 0.52 & 0.82 & 0.17 & 0\\
  5BZQ J2304--3625 & fsrq & 0.962 & 255 & J2304.9--3625 & 0.82 & 0.09 & 1.03 & 0.24 & 0\\
  5BZQ J2309--3059 & fsrq & 1.38 & 562 & J2309.2--3058 & 0.37 & 0.05 & --1.07 & 0.66 & 2\\
  5BZG J2310--4347 & bll-g & 0.0887 & 153 & J2310.7--4347 & 0.44 & 0.06 & 0.90 & 0.40 & 0\\
  5BZQ J2314--3138 & fsrq & 1.323 & 825 & J2314.8--3138 & 0.40 & 0.05 & 0.81 & 0.54 & 0\\
  5BZQ J2316--4041 & fsrq & 2.448 & 511 & J2316.8--4041 & 0.33 & 0.05 & 0.45 & 0.75 & 0\\
  5BZQ J2324--3714 & fsrq & 0.37 & 385 & J2324.1--3714 & 0.83 & 0.09 & 1.11 & 0.22 & 0\\
  5BZQ J2327--1447 & fsrq & 2.465 & 719 & J2327.7--1448 & 1.58 & 0.19 & 0.34 & 0.27 & 0\\
  5BZQ J2329--4730 & fsrq & 1.302 & 3180 & J2329.3--4730 & 2.79 & 0.30 & --0.03 & 0.18 & 0\\
  5BZQ J2329--4955 & fsrq & 0.518 & 558 & J2329.3--4955 & 0.58 & 0.07 & 0.97 & 0.29 & 0\\
  5BZQ J2330--4539 & fsrq & 0.447 & 1476 & J2330.6--4539 & 1.80 & 0.19 & --0.16 & 0.20 & 0\\
  5BZQ J2331--1556 & fsrq & 1.153 & 1335 & J2331.6--1557 & 2.31 & 0.25 & 0.55 & 0.20 & 0\\
  5BZU J2333--2343 & bcu & 0.0477 & 782 & J2333.9--2343 & 0.48 & 0.06 & 3.47 & 0.42 & 0\\
  5BZQ J2336--4115 & fsrq & 1.406 & 531 & J2336.5--4114 & 0.65 & 0.08 & --1.76 & 0.61 & 2\\
  5BZQ J2339--3310 & fsrq & 1.802 & 1284 & J2339.9--3310 & 3.62 & 0.39 & 0.54 & 0.18 & 0\\
  5BZQ J2343--2858 & fsrq & 1.936 & 112 & J2343.3--2858 & 0.26 & 0.04 & 0.68 & 0.51 & 0\\
  5BZQ J2348--1631 & fsrq & 0.576 & 2642 & J2348.0--1631 & 2.79 & 0.30 & 0.33 & 0.19 & 0\\
  5BZQ J2353--2743 & fsrq & 0.889 & 169 & J2353.1--2743 & 0.54 & 0.07 & 0.32 & 0.32 & 0\\
  5BZB J2353--3037 & bll & 0.737 & 397 & J2353.8--3038 & 0.56 & 0.07 & 0.62 & 0.32 & 0\\
  5BZU J2354--4106 & bcu & 0.632 & 620 & J2354.1--4106 & 1.09 & 0.12 & 0.57 & 0.23 & 0\\
  5BZQ J2354--1513 & fsrq & 2.675 & 865 & J2354.5--1513 & 1.16 & 0.14 & --0.49 & 0.34 & 0\\
  5BZQ J2355--3357 & fsrq & 0.702 & 330 & J2355.4--3358 & 0.98 & 0.11 & 1.29 & 0.24 & 0\\
  5BZQ J2357--5311 & fsrq & 1.006 & 1411 & J2357.9--5311 & 1.67 & 0.18 & 0.55 & 0.20 & 0\\
\hline
\end{longtable}
\end{longtab}

\clearpage

\setcounter{table}{5}

\begin{longtab}{}
\centering \small
\begin{longtable}{llrrrcrrlrrrr}%
\caption{\label{t.sample2} MWACS-3LAC sources}\\
\hline
  \multicolumn{1}{c}{3FGL name} &
  \multicolumn{1}{c}{Other name} &
  \multicolumn{1}{c}{$z$} &
  \multicolumn{1}{c}{Class} &
  \multicolumn{1}{c}{$S_{\gamma, E>0.1 \mathrm{GeV}}$} &
  \multicolumn{1}{c}{$\sigma_{S_\gamma}$} &
  \multicolumn{1}{c}{$\Gamma$} &
  \multicolumn{1}{c}{$\sigma_\Gamma$} &
  \multicolumn{1}{c}{MWACS name} &
  \multicolumn{1}{c}{$S_{0.18}$} &
  \multicolumn{1}{c}{$\sigma_{S_{0.18}}$} &
  \multicolumn{1}{c}{$\alpha_\mathrm{low}$} &
  \multicolumn{1}{c}{$\sigma_{\alpha_\mathrm{low}}$} \\
  \multicolumn{1}{c}{} &
  \multicolumn{1}{c}{} &
  \multicolumn{1}{c}{} &
  \multicolumn{1}{c}{} &
  \multicolumn{2}{c}{($10^{-12}$ erg cm$^{-2}$ s$^{-1}$)} &
  \multicolumn{1}{c}{} &
  \multicolumn{1}{c}{} &
  \multicolumn{1}{c}{} &
  \multicolumn{1}{c}{(Jy)} &
  \multicolumn{1}{c}{(Jy)} &
  \multicolumn{1}{c}{} &
  \multicolumn{1}{c}{} \\
\hline
  J0004.7--4740 & PKS 0002--478 & 0.88 & fsrq & 9.2 & 0.8 & 2.40 & 0.08 & J0004.5--4736 & 0.47 & 0.06 & 0.16 & 0.39\\
  J0030.3--4223 & PKS 0027--426 & 0.495 & fsrq & 19.8 & 0.9 & 2.58 & 0.04 & J0030.3--4224 & 0.69 & 0.08 & 0.01 & 0.29\\
  J0032.3--2852 & PMN J0032--2849 & 0.324 & bll & 2.6 & 0.6 & 2.19 & 0.18 & J0032.5--2849 & 0.40 & 0.06 & 0.20 & 0.47\\
  J0038.0--2501 & PKS 0035--252 & 1.196 & fsrq & 6.6 & 0.8 & 2.44 & 0.09 & J0038.2--2459 & 0.40 & 0.05 & --0.43 & 0.49\\
  J0039.0--2218 & PMN J0039--2220 & 0.06438 & bcu & 2.3 & 0.8 & 1.72 & 0.20 & J0039.1--2220 & 0.31 & 0.05 & 2.76 & 0.49\\
  J0049.4--5401 & PMN J0049--5402 & \dots & cand & 3.1 & 0.7 & 2.14 & 0.16 & J0049.8--5402 & 0.48 & 0.06 & 0.29 & 0.38\\
  J0049.8--5737 & PKS 0047--579 & 1.797 & fsrq & 6.1 & 0.8 & 2.46 & 0.11 & J0050.0--5738 & 3.41 & 0.37 & 0.77 & 0.18\\
  J0118.8--2142 & PKS 0116--219 & 1.165 & fsrq & 28.5 & 1.8 & 2.35 & 0.05 & J0118.9--2141 & 0.65 & 0.09 & 0.46 & 0.46\\
  J0120.4--2700 & PKS 0118--272 & \dots & bll & 42.4 & 2.2 & 1.91 & 0.03 & J0120.5--2701 & 1.80 & 0.19 & 0.46 & 0.19\\
  J0126.1--2227 & PKS 0123--226 & 0.72 & fsrq & 8.3 & 1.0 & 2.43 & 0.09 & J0126.2--2222 & 2.67 & 0.29 & 0.53 & 0.18\\
  J0132.6--1655 & PKS 0130--17 & 1.02 & fsrq & 26.2 & 1.1 & 2.43 & 0.04 & J0132.7--1655 & 1.30 & 0.15 & --0.27 & 0.28\\
  J0133.2--5159 & PKS 0131--522 & \dots & bcu & 3.3 & 0.7 & 2.63 & 0.19 & J0133.1--5200 & 0.31 & 0.05 & 0.46 & 0.64\\
  J0134.3--3842 & PMN J0134--3843 & 2.14 & fsrq & 3.9 & 0.7 & 2.33 & 0.14 & J0134.5--3843 & 0.94 & 0.11 & 0.68 & 0.27\\
  J0137.6--2430 & PKS 0135--247 & 0.835 & fsrq & 15.3 & 0.9 & 2.52 & 0.05 & J0137.6--2431 & 4.04 & 0.43 & 0.75 & 0.17\\
  J0145.1--2732 & PKS 0142--278 & 1.148 & fsrq & 21.8 & 1.0 & 2.57 & 0.04 & J0145.0--2733 & 0.95 & 0.11 & --0.06 & 0.31\\
  J0147.0--5204 & PKS 0144--522 & 0.0981 & bcu & 4.0 & 0.7 & 2.20 & 0.13 & J0146.8--5202 & 1.58 & 0.17 & 1.01 & 0.19\\
  J0205.2--1700 & PKS 0202--17 & 1.74 & fsrq & 18.2 & 0.9 & 2.76 & 0.05 & J0204.9--1701 & 1.15 & 0.14 & 0.08 & 0.31\\
  J0207.9--3846 & PKS 0205--391 & 0.254 & bcu & 4.3 & 0.7 & 2.54 & 0.14 & J0207.2--3857 & 0.95 & 0.11 & 0.44 & 0.24\\
  J0210.7--5101 & PKS 0208--512 & 1.003 & bcu & 51.1 & 1.5 & 2.17 & 0.03 & J0210.7--5100 & 6.02 & 0.64 & 0.50 & 0.17\\
  J0222.1--1616 & PKS 0219--164 & 0.698 & fsrq & 8.5 & 0.9 & 2.60 & 0.09 & J0222.0--1615 & 1.18 & 0.14 & 0.88 & 0.30\\
  J0228.3--5545 & PKS 0226--559 & 2.464 & fsrq & 12.2 & 0.8 & 2.43 & 0.06 & J0228.3--5545 & 0.34 & 0.06 & 0.92 & 0.77\\
  J0230.6--5757 & PKS 0229--581 & 0.03199 & bll & 2.3 & 0.7 & 1.68 & 0.29 & J0231.1--5754 & 0.38 & 0.07 & 1.73 & 0.60\\
  J0245.9--4651 & PKS 0244--470 & 1.385 & fsrq & 56.1 & 1.4 & 2.27 & 0.03 & J0246.0--4651 & 3.23 & 0.34 & 0.70 & 0.18\\
  J0252.8--2218 & PKS 0250--225 & 1.427 & fsrq & 57.1 & 1.6 & 2.15 & 0.03 & J0252.8--2219 & 0.94 & 0.11 & 0.35 & 0.27\\
  J0253.1--5438 & PKS 0252--549 & 0.539 & fsrq & 4.7 & 0.7 & 2.45 & 0.12 & J0253.5--5441 & 0.84 & 0.10 & 0.55 & 0.31\\
  J0303.4--2407 & PKS 0301--243 & 0.26 & bll & 65.9 & 2.7 & 1.92 & 0.02 & J0303.4--2407 & 2.50 & 0.27 & 0.45 & 0.18\\
  J0305.2--1607 & PKS 0302--16 & \dots & cand & 3.0 & 0.9 & 1.69 & 0.19 & J0305.2--1608 & 4.95 & 0.53 & 0.87 & 0.18\\
  J0326.0--1842 & PMN J0325--1843 & \dots & cand & 2.9 & 0.8 & 2.21 & 0.23 & J0325.9--1844 & 0.61 & 0.09 & --1.87 & 0.52\\
  J0334.3--4008 & PKS 0332--403 & 1.357 & bll & 42.3 & 1.7 & 2.00 & 0.04 & J0334.2--4008 & 0.44 & 0.07 & --0.41 & 0.66\\
  J0336.9--3622 & PKS 0335--364 & 1.537 & fsrq & 3.6 & 0.8 & 2.44 & 0.14 & J0336.9--3615 & 0.37 & 0.06 & 0.81 & 0.65\\
  J0339.2--1738 & PKS 0336--177 & 0.0655 & bcu & 6.8 & 1.1 & 1.93 & 0.10 & J0339.2--1736 & 0.52 & 0.08 & --0.17 & 0.61\\
  J0340.5--2119 & PKS 0338--214 & 0.223 & bll & 7.2 & 0.9 & 2.22 & 0.10 & J0340.6--2119 & 1.62 & 0.18 & 0.12 & 0.20\\
  J0343.2--2534 & PKS 0341--256 & 1.419 & fsrq & 9.4 & 0.9 & 2.56 & 0.08 & J0343.3--2530 & 1.45 & 0.16 & 0.82 & 0.19\\
  J0348.6--2748 & PKS 0346--27 & 0.991 & fsrq & 5.0 & 0.8 & 2.38 & 0.12 & J0348.6--2749 & 1.08 & 0.13 & --1.01 & 0.32\\
  J0357.1--4957 & PKS 0355--500 & 0.643 & bll & 4.3 & 0.7 & 2.10 & 0.13 & J0357.0--4955 & 0.44 & 0.06 & 1.64 & 0.40\\
  J0359.3--2612 & PKS 0357--264 & \dots & bll & 2.9 & 0.8 & 2.19 & 0.18 & J0359.5--2615 & 2.40 & 0.26 & 0.10 & 0.19\\
  J0401.8--3144 & PKS 0400--319 & 1.288 & fsrq & 3.6 & 0.7 & 2.54 & 0.20 & J0402.3--3147 & 0.37 & 0.05 & --0.59 & 0.56\\
  J0403.7--2442 & TXS 0401--248 & 0.598 & fsrq & 2.9 & 0.7 & 2.37 & 0.18 & J0403.7--2444 & 0.29 & 0.05 & 2.86 & 0.47\\
  J0403.9--3604 & PKS 0402--362 & 1.417 & fsrq & 85.3 & 1.5 & 2.27 & 0.03 & J0403.9--3605 & 1.48 & 0.16 & --0.11 & 0.19\\
  J0407.1--3825 & PKS 0405--385 & 1.285 & fsrq & 21.7 & 1.1 & 2.40 & 0.04 & J0407.0--3826 & 0.46 & 0.07 & --0.50 & 0.62\\
  J0416.6--1850 & PKS 0414--189 & 1.421 & fsrq & 14.0 & 1.0 & 2.34 & 0.06 & J0416.6--1851 & 0.45 & 0.08 & --0.15 & 0.63\\
  J0425.0--5331 & PMN J0425--5331 & 0.39 & bll & 9.3 & 0.8 & 2.34 & 0.08 & J0425.1--5331 & 0.40 & 0.06 & 1.00 & 0.53\\
  J0428.6--3756 & PKS 0426--380 & 1.105 & bll & 184.0 & 4.2 & 1.95 & 0.02 & J0428.6--3756 & 0.92 & 0.11 & 0.77 & 0.24\\
  J0449.4--4350 & PKS 0447--439 & 0.205 & bll & 123.1 & 3.8 & 1.85 & 0.02 & J0449.4--4350 & 0.99 & 0.11 & 0.49 & 0.24\\
  J0453.2--2808 & PKS 0451--28 & 2.564 & fsrq & 25.0 & 1.0 & 2.63 & 0.04 & J0453.2--2807 & 2.63 & 0.29 & --0.11 & 0.18\\
  J0455.7--4617 & PKS 0454--46 & 0.858 & fsrq & 22.8 & 1.0 & 2.55 & 0.04 & J0455.8--4615 & 7.23 & 0.77 & 0.70 & 0.17\\
  J0505.5--1558 & TXS 0503--160 & \dots & cand & 4.1 & 0.8 & 2.10 & 0.14 & J0505.6--1558 & 1.94 & 0.22 & 0.53 & 0.23\\
  J0521.4--1740 & TXS 0519--176 & 0.347 & fsrq & 6.4 & 0.9 & 2.43 & 0.11 & J0521.3--1737 & 1.21 & 0.14 & 0.44 & 0.26\\
  J0522.9--3628 & PKS 0521--36 & 0.0553 & bcu & 61.8 & 1.6 & 2.34 & 0.03 & J0522.9--3627 & 61.86 & 6.39 & 0.62 & 0.14\\
  J0525.3--4558 & PKS 0524--460 & 1.479 & fsrq & 3.6 & 0.8 & 2.20 & 0.16 & J0525.5--4557 & 3.85 & 0.41 & 0.91 & 0.18\\
  J0525.8--2014 & PMN J0525--2010 & \dots & cand & 3.6 & 0.8 & 2.06 & 0.16 & J0525.4--2011 & 1.14 & 0.14 & 0.95 & 0.30\\
  J0526.2--4829 & PKS 0524--485 & 1.299 & fsrq & 17.1 & 1.3 & 2.29 & 0.06 & J0526.3--4830 & 0.54 & 0.07 & 0.60 & 0.43\\
  J0538.8--4405 & PKS 0537--441 & 0.892 & bll & 308.3 & 4.9 & 1.93 & 0.01 & J0538.8--4405 & 3.50 & 0.37 & --0.29 & 0.19\\
  J0540.0--2837 & PKS 0537--286 & 3.104 & fsrq & 17.3 & 1.1 & 2.78 & 0.06 & J0539.9--2839 & 0.58 & 0.07 & --0.66 & 0.43\\
  J0540.5--5416 & PKS 0539--543 & 1.185 & fsrq & 12.0 & 0.9 & 2.63 & 0.07 & J0540.7--5418 & 0.69 & 0.09 & 0.81 & 0.39\\
  J0550.6--3217 & PKS 0548--322 & 0.069 & bll & 4.0 & 1.0 & 1.61 & 0.16 & J0550.6--3216 & 2.32 & 0.25 & 0.75 & 0.19\\
  J0600.9--3943 & PKS 0558--396 & 1.661 & fsrq & 6.6 & 1.2 & 2.81 & 0.13 & J0600.5--3937 & 0.64 & 0.09 & --0.41 & 0.54\\
  J0629.4--1959 & PKS 0627--199 & 1.724 & bll & 38.2 & 1.8 & 2.18 & 0.03 & J0629.4--1959 & 0.67 & 0.10 & --0.10 & 0.55\\
  J0648.1--3045 & PKS 0646--306 & 1.153 & fsrq & 15.7 & 1.2 & 2.56 & 0.07 & J0648.2--3044 & 1.25 & 0.14 & 0.39 & 0.24\\
  J0706.1--4849 & PMN J0705--4847 & \dots & cand & 9.0 & 1.2 & 2.41 & 0.09 & J0705.9--4847 & 0.33 & 0.05 & 1.17 & 0.55\\
  J0726.6--4727 & PMN J0726--4728 & 1.686 & fsrq & 17.4 & 1.4 & 2.46 & 0.06 & J0726.4--4728 & 0.40 & 0.07 & 0.32 & 0.71\\
  J0732.2--4638 & PKS 0731--465 & \dots & cand & 9.3 & 1.2 & 2.35 & 0.17 & J0732.7--4640 & 2.08 & 0.23 & 0.21 & 0.21\\
  J2051.8--5535 & PMN J2052--5533 & \dots & cand & 6.0 & 1.0 & 2.58 & 0.15 & J2052.2--5532 & 0.71 & 0.11 & 0.65 & 0.53\\
  J2056.2--4714 & PKS 2052--47 & 1.491 & fsrq & 77.4 & 1.8 & 2.26 & 0.03 & J2056.2--4714 & 4.50 & 0.48 & 0.86 & 0.18\\
  J2126.5--4605 & PKS 2123--463 & 1.67 & fsrq & 28.7 & 1.1 & 2.50 & 0.04 & J2126.5--4605 & 2.84 & 0.30 & 0.83 & 0.18\\
  J2135.3--5008 & PMN J2135--5006 & 2.181 & fsrq & 12.0 & 1.2 & 2.47 & 0.07 & J2135.3--5006 & 0.17 & 0.03 & 3.02 & 0.62\\
  J2141.7--3734 & PKS 2138--377 & 0.423 & fsrq & 6.7 & 0.9 & 2.59 & 0.11 & J2141.8--3728 & 0.85 & 0.10 & 0.44 & 0.36\\
  J2143.1--3928 & PMN J2143--3929 & 0.429 & bll & 5.7 & 0.9 & 2.07 & 0.13 & J2143.0--3928 & 0.37 & 0.05 & 1.31 & 0.45\\
  J2151.6--2744 & PMN J2151--2742 & 1.485 & fsrq & 4.4 & 0.8 & 2.51 & 0.15 & J2151.3--2742 & 0.41 & 0.06 & 1.45 & 0.57\\
  J2151.8--3025 & PKS 2149--306 & 2.345 & fsrq & 29.6 & 1.3 & 2.61 & 0.08 & J2151.9--3027 & 0.82 & 0.10 & 1.50 & 0.46\\
  J2158.0--1501 & PKS 2155--152 & 0.672 & fsrq & 12.6 & 1.1 & 2.27 & 0.07 & J2158.0--1501 & 2.48 & 0.28 & 0.53 & 0.25\\
  J2158.8--3013 & PKS 2155--304 & 0.116 & bll & 230.5 & 6.2 & 1.75 & 0.02 & J2158.8--3013 & 0.82 & 0.10 & 0.92 & 0.33\\
  J2207.8--5345 & PKS 2204--54 & 1.215 & fsrq & 12.9 & 0.8 & 2.58 & 0.06 & J2207.7--5346 & 2.91 & 0.31 & 0.74 & 0.18\\
  J2213.1--2532 & PKS 2210--25 & 1.833 & fsrq & 7.6 & 0.8 & 2.55 & 0.10 & J2213.0--2529 & 2.46 & 0.26 & 0.73 & 0.19\\
  J2222.3--3500 & PKS 2220--351 & 0.298 & fsrq & 3.1 & 0.6 & 2.37 & 0.18 & J2223.1--3455 & 2.86 & 0.31 & 0.52 & 0.18\\
  J2230.6--4419 & PKS 2227--445 & 1.326 & fsrq & 3.7 & 0.7 & 2.57 & 0.17 & J2230.9--4416 & 0.51 & 0.06 & 0.96 & 0.30\\
  J2243.4--2541 & PKS 2240--260 & 0.774 & bll & 19.6 & 1.1 & 2.27 & 0.05 & J2243.4--2544 & 2.63 & 0.28 & 0.32 & 0.18\\
  J2248.6--3235 & PKS 2245--328 & 2.268 & fsrq & 5.0 & 0.7 & 2.82 & 0.14 & J2248.6--3235 & 0.39 & 0.05 & --2.40 & 0.48\\
  J2250.7--2806 & PMN J2250--2806 & 0.525 & bcu & 47.1 & 1.7 & 2.18 & 0.03 & J2250.7--2806 & 0.26 & 0.04 & 1.07 & 0.51\\
  J2258.0--2759 & PKS 2255--282 & 0.926 & fsrq & 48.7 & 1.5 & 2.17 & 0.04 & J2258.1--2758 & 1.36 & 0.15 & 0.58 & 0.20\\
  J2329.3--4955 & PKS 2326--502 & 0.518 & fsrq & 147.6 & 2.3 & 2.12 & 0.02 & J2329.3--4955 & 0.58 & 0.07 & 0.97 & 0.29\\
  J2329.9--4734 & PKS 2326--477 & 1.302 & fsrq & 4.3 & 0.9 & 2.23 & 0.16 & J2329.3--4730 & 2.79 & 0.30 & --0.03 & 0.18\\
  J2336.5--4116 & PKS 2333--415 & 1.406 & fsrq & 11.3 & 1.0 & 2.24 & 0.07 & J2336.5--4114 & 0.65 & 0.08 & --1.76 & 0.61\\
  J2348.0--1630 & PKS 2345--16 & 0.576 & fsrq & 29.4 & 1.7 & 2.20 & 0.05 & J2348.0--1631 & 2.79 & 0.30 & 0.33 & 0.19\\
  J2353.6--3037 & PKS 2351--309 & 0.737 & bll & 5.3 & 0.9 & 2.28 & 0.13 & J2353.8--3038 & 0.56 & 0.07 & 0.62 & 0.32\\
  J2357.8--5310 & PKS 2355--534 & 1.006 & fsrq & 11.8 & 1.2 & 2.46 & 0.07 & J2357.9--5311 & 1.67 & 0.18 & 0.55 & 0.20\\
  J2359.5--2052 & TXS 2356--210 & 0.096 & bll & 3.8 & 0.7 & 2.02 & 0.16 & J2359.3--2048 & 1.37 & 0.15 & 0.53 & 0.21\\
\hline
\end{longtable}
\end{longtab}

\clearpage

\setcounter{table}{6}

\begin{table*}
\centering
\caption{\label{t.sample3} UGS with MWACS sources in the 95\% error ellipse}
\begin{tabular}{lrllllllllrr}
\hline
  \multicolumn{1}{c}{3FGL name} &
  \multicolumn{1}{c}{$S_{\gamma, E>0.1 \mathrm{GeV}}$} &
  \multicolumn{1}{c}{$\sigma_{S_\gamma}$} &
  \multicolumn{1}{c}{$\Gamma$} &
  \multicolumn{1}{c}{$\sigma_\Gamma$} &
  \multicolumn{1}{c}{MWACS name} &
  \multicolumn{1}{c}{$S_{0.18}$} &
  \multicolumn{1}{c}{$\sigma_{S_{0.18}}$} &
  \multicolumn{1}{c}{$\alpha_\mathrm{low}$} &
  \multicolumn{1}{c}{$\sigma_{\alpha_\mathrm{low}}$} \\
  \multicolumn{1}{c}{} &
  \multicolumn{2}{c}{($10^{-12}$\eflux)} &
  \multicolumn{1}{c}{} &
  \multicolumn{1}{c}{} &
  \multicolumn{1}{c}{} &
  \multicolumn{1}{c}{(Jy)} &
  \multicolumn{1}{c}{(Jy)} &
  \multicolumn{1}{c}{} &
  \multicolumn{1}{c}{} \\
\hline
  3FGL J0026.2--4812 & 3.58 & 0.68 & 2.19 & 0.15 & J0025.6--4816 & 0.65 & 0.08 & 0.18 & 0.30\\
  3FGL J0031.2--2320 & 2.66 & 0.66 & 2.14 & 0.17 & J0030.6--2323 & 0.26 & 0.04 & --0.26 & 0.80\\
  3FGL J0121.8--3917 & 2.43 & 0.69 & 1.80 & 0.20 & J0122.1--3919 & 0.32 & 0.05 & 0.50 & 0.43\\
  3FGL J0133.2--4737 & 3.08 & 0.68 & 2.71 & 0.20 & J0132.2--4742 & 0.32 & 0.05 & --0.97 & 0.74\\
  3FGL J0226.7--4747 & 4.60 & 0.75 & 2.92 & 0.19 & J0226.0--4744 & 0.80 & 0.09 & 1.40 & 0.22\\
   & & & & & J0226.3--4750 & 0.18 & 0.03 & 1.40 & 0.22\\
  3FGL J0308.4--2852 & 2.98 & 0.72 & 2.54 & 0.25 & J0308.2--2852 & 1.25 & 0.14 & 0.84 & 0.21\\
  3FGL J0608.2--2306 & 3.27 & 0.78 & 2.20 & 0.16 & J0607.5--2308 & 0.34 & 0.06 & 1.04 & 0.65\\
  3FGL J0714.7--3924 & 5.64 & 1.26 & 2.38 & 0.14 & J0714.1--3923 & 0.36 & 0.05 & 0.90 & 0.55\\
  3FGL J0718.9--5004 & 5.07 & 1.00 & 2.39 & 0.14 & J0719.2--5015 & 0.37 & 0.05 & 0.31 & 0.51\\
   & & & & & J0719.3--4955 & 0.43 & 0.06 & 1.01 & 0.41\\
  3FGL J2130.4--4237 & 8.01 & 0.97 & 2.56 & 0.10 & J2130.2--4243 & 0.64 & 0.08 & 0.69 & 0.36\\
   &  & & & & J2131.1--4234 & 1.04 & 0.12 & 0.54 & 0.25\\
\hline\end{tabular}
\end{table*}


\end{document}